\documentclass[lettersize,journal]{IEEEtran}
\usepackage{amsmath,amsfonts}
\usepackage{algorithmic}
\usepackage{algorithm}
\usepackage{array}
\usepackage[caption=false,font=footnotesize,labelfont=rm,textfont=rm]{subfig}
\usepackage{textcomp}
\usepackage{stfloats}
\usepackage{url}
\usepackage{verbatim}
\usepackage{graphicx}
\usepackage{color,xcolor}
\usepackage{cite}
\usepackage{subfloat}
\usepackage{makecell}
\usepackage{multirow}
\usepackage{threeparttable, booktabs}

\begin{document}

\title{Near-Field Propagation and Spatial Non-Stationarity Channel Model for 6-24 GHz (FR3) Extremely Large-Scale MIMO: Adopted by 3GPP for 6G}

\author{Huixin Xu, Jianhua Zhang, Pan Tang, Hongbo Xing, Haiyang Miao, Nan Zhang, Jian Li, Jianming Wu, Wenfei Yang, Zhening Zhang, Wei Jiang, Zijian He, Afshin Haghighat, Qixing Wang, and Guangyi Liu

\thanks{This research is supported in part by National Natural Science Foundation of China under Grant 62525101, Grant 62201086, in part by the National Key R\&D Program of China under Grant 2023YFB2904805, in part by the Guangdong Major Project of Basic and Applied Basic Research under Grant 2023B0303000001, and in part by the Beijing University of Posts and Telecommunications-China Mobile Research Institute Joint Innovation Center. \textit{(Corresponding author: Pan Tang.)}}
\thanks{Huixin Xu, Jianhua Zhang, Pan Tang, Hongbo Xing, and Haiyang Miao are with the State Key Lab of Networking and Switching Technology, Beijing University of Posts and Telecommunications, Beijing 100876, China (e-mail: xuhuixin@bupt.edu.cn; jhzhang@bupt.edu.cn; tangpan27@bupt.edu.cn; hbxing@bupt.edu.cn; hymiao@bupt.edu.cn).}
\thanks{Nan Zhang is with ZTE Corporation, Shanghai 201203, China (e-mail: zhang.nan152@zte.com.cn).}
\thanks{Jian Li, Wenfei Yang, and Zhening Zhang are with the Wireless Technology Lab, Huawei Technologies Co., Ltd., Shanghai 201206, China (e-mail: Calvin.li@huawei.com; yangwenfei4@huawei.com; zhangzhening1@huawei.com).}
\thanks{Jianming Wu, Wei Jiang, and Zijian He are with vivo Mobile Communication Co., Ltd., Dongguan 523850, China (e-mail: jianming.wu@vivo.com; wei.jiang@vivo.com; zijian.he@vivo.com).}
\thanks{Afshin Haghighat is with InterDigital Canada, Montreal, QC H3A 3G4, Canada (e-mail: afshin.haghighat@interdigital.com).}
\thanks{Qixing Wang and Guangyi Liu are with China Mobile Research Institute, Beijing 100053, China (e-mail: wangqixing@chinamobile.com; liuguangyi@chinamobile.com).}
}

\markboth{Journal of \LaTeX\ Class Files,~Vol.~14, No.~8, August~2021}%
{Shell \MakeLowercase{\textit{et al.}}: A Sample Article Using IEEEtran.cls for IEEE Journals}

\maketitle

\begin{abstract}

Next generation cellular deployments are expected to exploit the 6–24 GHz frequency range 3 (FR3) and extremely large-scale multiple-input multiple-output (XL-MIMO) to enable ultra-high data rates and reliability. However, the significantly enlarged antenna apertures and higher carrier frequencies render the far-field and spatial stationarity assumptions in the existing 3rd generation partnership project (3GPP) channel models invalid, giving rise to new features such as near-field propagation and spatial non-stationarity (SNS). Despite extensive prior research, incorporating these new features within the standardized channel modeling framework remains an open issue. To address this, this paper presents a channel modeling framework for XL-MIMO systems that incorporates both near-field and SNS features, adopted by 3GPP. For the near-field propagation feature, the framework models the distances from the base station (BS) and user equipment to the spherical-wave sources associated with clusters. These distances are used to characterize element-wise variations of path parameters, such as nonlinear changes in phase and angle. To capture the effect of SNS at the BS side, a stochastic-based approach is proposed to model SNS caused by incomplete scattering, by establishing power attenuation factors from visibility probability and visibility region to characterize antenna element-wise path power variation. In addition, a physical blocker-based approach is introduced to model SNS effects caused by partial blockage. Finally, a simulation framework for near-field and SNS is developed within the structure of the existing 3GPP channel model. Performance evaluations demonstrate that the near-field model captures higher channel capacity potential compared to the far-field model. Coupling loss results indicate that SNS leads to more pronounced propagation fading relative to the spatial stationary model.
\end{abstract}


\section{Introduction}
\IEEEPARstart{S}{ixth}-generation (6G) wireless networks are envisioned to enable significantly higher data rates, along with ubiquitous connectivity and hyper reliability \cite{C1_6G_survey,DTC,6G,C1_FR3_MHY}. Achieving these ambitious goals necessitates access to new spectrum bands that offer both wide bandwidth and reliable coverage. The 6–24 GHz range, known as frequency range 3 (FR3) band, is considered particularly promising as it offers an attractive balance between large contiguous bandwidth and relatively low propagation loss. To further boost spectral efficiency and spatial resolution, extremely large-scale multiple-input multiple-output (XL-MIMO) systems—with hundreds or even thousands of antennas at the base station (BS)—have emerged as a key technology for 6G \cite{XL_MIMO}. Therefore, the integration of XL-MIMO and the FR3 band is a promising solution to meet 6G’s dual demands for capacity and coverage.

However, the deployment of XL-MIMO in the FR3 band introduces fundamental challenges to conventional channel modeling assumptions, particularly the far-field propagation and spatial stationarity (SS). The combination of large antenna apertures and shorter wavelengths increases the likelihood that user equipment (UE) operates within the near-field region, where the errors of the conventional plane-wave assumption become non-negligible, necessitating the adoption of spherical-wavefront models for accurate characterization \cite{TP_XLMIMO}. Moreover, the enlarged array size introduces spatial non-stationarity (SNS), characterized by antenna element-wise path power variation. The underlying causes of SNS can be categorized into two types. First, objects with limited size may not fully block the entire array, resulting in \textit{partial blockage}. Second, objects with limited size may no longer act as complete scatterers for all antenna elements, leading to \textit{incomplete scattering} \cite{3GPP_vivo_helan,C1_WSSUS}.

Extensive channel measurement and modeling studies have been conducted on near-field propagation and SNS in XL-MIMO systems. For near-field modeling, one approach divides the large antenna array into subarrays, which applies a planar wave approximation within each subarray and a spherical-wave model among subarrays \cite{C1_Hybrid_HC,C1_MIMO_MHY}. Another approach extends existing 3GPP standard models by incorporating the positions of scatterers to capture the spherical-wave effects \cite{C1_MIMO_yxf,C1_NF_HC,Quadriga}. However, these approaches remain inadequate in ensuring consistency between near-field and far-field, and scatterer locations are difficult to determine. In terms of SNS modeling, two mainstream approaches have been proposed: visibility region (VR)-based model and birth-death process. In VR-based models, each cluster is associated with a specific VR of the array \cite{C1_VR_COST,C1_VR_ljz}. In contrast, the birth-death process models SNS by capturing the evolution of clusters along the array dimension, where their birth and death are treated as a stochastic process  \cite{C1_cluster_LR,C1_BD_CX}. Moreover, a simplified model introduces an additional parameter to characterize SNS-induced power variations resulting from blockage, reflection, and diffraction \cite{C3_YZQ,C1_THz_MIMO_xhx}. Nevertheless, these models have yet to provide effective solutions for determining the size of VRs or birth-death probabilities. In summary, existing modeling approaches for near-field propagation and SNS cannot be directly extended to standardized channel models due to fundamental compatibility and parameterization challenges.

The 3rd generation partnership project (3GPP) technical specification group (TSG) for radio access network (RAN) approved a new study item on channel modeling enhancements for 7-24 GHz during the RAN \#102 meeting. One key objective was to adapt and extend TR 38.901 to account for FR3-specific propagation characteristics, especially those arising from XL-MIMO deployments. These include near-field propagation and SNS effects—the modeling of ray cluster blockages and/or channel parameter correlation effect on a subset of the antenna elements of a large antenna array \cite{3GPP_SID,C1_TP}. This study was conducted by RAN Working Group 1 (RAN1), with technical discussions initiated at the RAN1 \#116-bis meeting and concluded at the RAN1 \#121, spanning eight formal meetings in total. Ultimately, new channel modeling features supporting near-field propagation and SNS were formally defined and are being incorporated into the 3GPP TR 38.901 technical report. This paper presents a unified near-field propagation and SNS channel modeling framework for FR3 XL-MIMO systems adopted by 3GPP. The key contributions are summarized as follows:
\begin{itemize}

\item Empirical observations reveal nonlinear phase variations and angle shifts across the array due to near-field propagation, as well as path power variations caused by SNS from partial blockage and incomplete scattering. To model these effects, we present a unified channel modeling framework that jointly captures near-field and SNS features for XL-MIMO systems.

\item To characterize the near-field feature, the distances from the BS and UE to the spherical-wave sources associated with clusters are introduced, along with their generation method. Specifically, for the $N_{\rm{spec}}$ strongest clusters, the distance equals the total propagation path length. For other clusters, it is obtained by scaling the absolute path length with a factor drawn from a Beta distribution.

\item To characterize the SNS feature, two alternative approaches are introduced at the BS side. The stochastic-based approach uses visibility probability (VP) to define VR and generate power attenuation factors across antenna elements. The physical blocker-based approach, aligned with 3GPP Blockage Model B, computes element-wise power variations caused by obstructions. At the UE side, power attenuation is modeled by assigning fixed attenuation values to each antenna element.

\item The XL-MIMO near-field propagation and SNS channel simulation framework is developed within the existing 3GPP modeling structure. Comparative evaluations against conventional far-field and SS models—using metrics such as channel capacity and coupling loss—demonstrate the effectiveness of the adopted framework in capturing near-field and SNS features.
\end{itemize}

This paper is organized as follows. Section II presents the evolution from MIMO to XL-MIMO channel modeling framework, incorporating near-field propagation and SNS. Sections III and IV detail the modeling approaches for near-field propagation and SNS features, respectively. Section V presents the XL-MIMO channel simulation framework and evaluates the model performance. Finally, Section VI concludes the paper.

\section{From MIMO to XL-MIMO: Evolution of Channel Modeling Frameworks}
\label{Sec_2}

In this section, we first introduce the MIMO channel modeling framework standardized in 3GPP TR 38.901 for 5G systems. However, empirical observations reveal that this framework does not capture two essential features—near-field propagation and SNS—which are prominent in XL-MIMO systems. To overcome these limitations, a new channel modeling framework tailored for XL-MIMO is presented.

\subsection{MIMO Channel Modeling Framework}

The channel impulse response (CIR) between the $u$-th receive (RX) and $s$-th transmit (TX) antenna elements, denoted as $H_{u,s}(\tau, t)$, is modeled as the superposition of line-of-sight (LOS) and non-line-of-sight (NLOS) components, scaled by a specified Ricean K-factor, as follows:
\begin{equation}
\begin{aligned}
H_{u,s}^{{\rm{LOS}}}\left( {\tau ,t} \right) =& \sqrt {\frac{1}{{{K_R} + 1}}} H_{u,s}^{{\rm{NLOS}}}\left( {\tau ,t} \right) + \\
&\sqrt {\frac{{{K_R}}}{{{K_R} + 1}}} H_{u,s,1}^{{\rm{LOS}}}\left( t \right)\delta \left( {\tau  - {\tau _1}} \right),
\end{aligned} 
\label{equ_CIR_LOS}
\end{equation}
where \(K_R\) is the Ricean K-factor in linear scale. When the Ricean K-factor $K_R = 0$, the channel reduces to a purely NLOS condition. $H_{u,s,1}^{\mathrm{LOS}}(t)$ denotes the channel coefficient of the LOS component (see Eq.~(\ref{equ_H_LOS_5G})), while $H_{u,s}^{\mathrm{NLOS}}(\tau, t)$ represents the CIR of the NLOS component. 
\begin{figure*}
\begin{equation}
\begin{aligned}
H_{u,s,1}^{{\rm{LOS}}}\left( t \right) = & \left[ {\begin{array}{*{20}{c}}
{{F_{rx,u,\theta }}\left( {{\theta _{{\rm{LOS}},ZOA}},{\phi _{{\rm{LOS}},AOA}}} \right)}\\
{{F_{rx,u,\phi }}\left( {{\theta _{{\rm{LOS}},ZOA}},{\phi _{{\rm{LOS}},AOA}}} \right)}
\end{array}} \right]\left[ {\begin{array}{*{20}{c}}
1&0\\
0&{ - 1}
\end{array}} \right]\left[ {\begin{array}{*{20}{c}}
{{F_{tx,s,\theta }}\left( {{\theta _{{\rm{LOS}},ZOD}},{\phi _{{\rm{LOS}},AOD}}} \right)}\\
{{F_{tx,s,\phi }}\left( {{\theta _{{\rm{LOS}},ZOD}},{\phi _{{\rm{LOS}},AOD}}} \right)}
\end{array}} \right] \\
& \cdot \exp \left( { - j2\pi \frac{{{d_{{\rm{3D}}}}}}{{{\lambda _0}}}} \right)\exp \left( {j2\pi \frac{{\hat r_{rx,{\rm{LOS}}}^T{{\bar d}_{rx,u}}}}{{{\lambda _0}}}} \right)\exp \left( {j2\pi \frac{{\hat r_{tx,{\rm{LOS}}}^T{{\bar d}_{tx,s}}}}{{{\lambda _0}}}} \right)\exp \left( {j2\pi \frac{{\hat r_{rx,{\rm{LOS}}}^T\bar v}}{{{\lambda _0}}}t} \right).
\end{aligned} 
\label{equ_H_LOS_5G}
\end{equation}
\end{figure*}
The NLOS component consists of \(N\) clusters, each containing \(M\) rays, and is given by:
\begin{equation}
\begin{aligned}
H_{u,s}^{{\rm{NLOS}}}\left( {\tau ,t} \right) = &\sum\limits_{n = 1}^2 {\sum\limits_{i = 1}^3 {\sum\limits_{m \in {R_i}} {H_{u,s,n,m}^{{\rm{NLOS}}}(t)\delta \left( {\tau  - {\tau _{n,i}}} \right) } } } \\
&+ \sum\limits_{n = 3}^N {\sum\limits_{m = 1}^M {H_{u,s,n,m}^{{\rm{NLOS}}}(t)\delta \left( {\tau  - {\tau _n}} \right)}}.
\end{aligned} 
\label{equ_CIR_NLOS}
\end{equation}
For the two strongest clusters (i.e., $n = 1,2$), rays are spread in delay to three sub-clusters (per cluster), with fixed delay offset indexed by $i$. The NLOS channel coefficient \(H_{u,s,n,m}^{{\rm{NLOS}}}(t)\) of the $m$-th ray in the $n$-th cluster is defined in Eq.~(\ref{equ_H_NLOS_5G}). The associated small-scale parameters include the normalized power $P_n$, the relative path delays $\tau_n$ or $\tau_{n,i}$, and the polarization matrix $\mathbf{\Phi}_{n,m}$, given by $\mathbf{\Phi}_{n,m} =
\begin{bmatrix}
e^{j \Phi_{n,m}^{\theta\theta}} & \sqrt{\kappa_{n,m}^{-1}} e^{j \Phi_{n,m}^{\theta\phi}} \\
\sqrt{\kappa_{n,m}^{-1}} e^{j \Phi_{n,m}^{\phi\theta}} & e^{j \Phi_{n,m}^{\phi\phi}}
\end{bmatrix}$. Here, $\kappa_{n,m}$ denotes the cross-polarization power ratio (XPR), and $\Phi_{n,m}^{\theta\theta}, \Phi_{n,m}^{\theta\phi}, \Phi_{n,m}^{\phi\theta}, \Phi_{n,m}^{\phi\phi}$ are independent random phases uniformly distributed over $[0, 2\pi)$. Other parameters such as angular-domain variables, antenna field patterns, and spherical unit vectors are summarized in Table~\ref{Tab_Par}. In addition, $\lambda_0$ denotes the carrier wavelength, and $\bar{v}$ represents the velocity vector of the UE.

\begin{figure*}
\begin{equation}
\begin{aligned}
H_{u,s,n,m}^{\rm{NLOS}}\left( t \right) = &\sqrt {\frac{{{P_n}}}{M}} \left[ {\begin{array}{*{20}{c}}
{{F_{rx,u,\theta }}\left( {{\theta _{n,m,{\rm{ZOA}}}},{\phi _{n,m,{\rm{AOA}}}}} \right)}\\
{{F_{rx,u,\phi }}\left( {{\theta _{n,m,{\rm{ZOA}}}},{\phi _{n,m,{\rm{AOA}}}}} \right)}
\end{array}} \right] \mathbf{\Phi}_{n,m}  \left[ {\begin{array}{*{20}{c}}
{{F_{tx,s,\theta }}\left( {{\theta _{n,m,{\rm{ZOD}}}},{\phi _{n,m,{\rm{AOD}}}}} \right)}\\
{{F_{tx,s,\phi }}\left( {{\theta _{n,m,{\rm{ZOD}}}},{\phi _{n,m,{\rm{AOD}}}}} \right)}
\end{array}} \right] \\ 
& \cdot\exp \left( {j2\pi \frac{{\hat r_{rx,n,m}^T{{\bar d}_{rx,u}}}}{{{\lambda _0}}}} \right)\exp \left( {j2\pi \frac{{\hat r_{tx,n,m}^T{{\bar d}_{tx,s}}}}{{{\lambda _0}}}} \right)\exp \left( {j2\pi \frac{{\hat r_{rx,n,m}^T\bar v}}{{{\lambda _0}}}t} \right).
\end{aligned} 
\label{equ_H_NLOS_5G}
\end{equation}
\end{figure*}

\begin{table*}[]
\renewcommand\arraystretch{1.2}
    \caption{Definitions of key parameters in MIMO and XL-MIMO channel models}
    \centering
    \begin{tabular}{m{3cm}<{\centering}|m{3.8cm}<{\centering}|m{10cm}<{\centering}}
    \hline
   MIMO Channel Model Symbols & XL-MIMO Channel Model Symbols  & Definitions  \\
    \hline
\vspace{7pt} \(\phi_{\rm{LOS},AOD}\), \(\phi_{\rm{LOS},AOA}\), \(\theta_{\rm{LOS},ZOD}\), \(\theta_{\rm{LOS},ZOA}\) & \vspace{7pt}  \(\phi_{{\rm{LOS},AOD,}u,s}\), \(\phi_{{\rm{LOS},AOA,}u,s}\),  \(\theta_{{\rm{LOS},ZOD,}u,s}\), \(\theta_{{\rm{LOS,ZOA}},u,s}\) & {\begin{tabular}{c} Far-field: LOS azimuth angle of departure (AOD), azimuth angle of arrival (AOA), \\zenith angle of departure (ZOD), and zenith angle of arrival (ZOA); \\ Near-field: AOD, AOA, ZOD, and ZOA of the TX–RX \((s, u)\) pair\end{tabular}}\\
\hline
\vspace{6pt} \(\phi_{n,m,{\rm{AOD}}}\), \(\phi_{n,m,{\rm{AOA}}}\), \(\theta_{n,m,{\rm{ZOD}}}\), \(\theta_{n,m,{\rm{ZOA}}}\)& \vspace{6pt} \(\phi_{n,m,{\rm{AOD}},s}\), \(\phi_{n,m,{\rm{AOA}},u}\), \(\theta_{n,m,{\rm{ZOD}},s}\), \(\theta_{n,m,{\rm{ZOA}},u}\) &\makecell{Far-field: AOD, AOA, ZOD, and ZOA for ray \(m\) in cluster \(n\); \\ Near-field: AOD, ZOD (TX element \(s\)) and AOA, ZOA (RX element \(u\)) \\ for ray \(m\) in cluster \(n\)}\\
    \hline
   \raisebox{-1.3ex}{\({{{\hat r}_{rx,{\rm{LOS}}}}}\) ,\({{{\hat r}_{tx,{\rm{LOS}}}}}\)}  & \raisebox{-1.3ex}{\({{{\vec r}_{u,s}}}\)} & \makecell*[c]{ Far-field: Spherical unit vector of LOS path for RX and TX, respectively \\ Near-field: Vector from TX element \(s\) to RX element \(u\)} \\ 
    \hline
    - & \(d_{1,n,m}\)  & Distance from BS to the spherical-wave source for ray \(m\) in cluster \(n\) \\
    \hline
    - &  \(d_{2,n,m}\) & Distance from UE to the spherical-wave source for ray \(m\) in cluster \(n\) \\
    \hline
    - & \(\alpha_{s,n,m}\)  & Power attenuation factor for ray \(m\) in cluster \(n\) at TX antenna element \(s\)\\
    \hline
    - &  \(\beta_{u}\) & Power attenuation factor at RX antenna element \(u\) \\
    \hline
    \multicolumn{2}{c|}{\({F_{rx,u,\theta }}\), \({F_{rx,u,\phi }}\) }&   Field patterns of RX element \(u\) along spherical basis vectors \(\hat{\theta}\) and \(\hat{\phi}\), respectively \\
    \hline
    \multicolumn{2}{c|}{\({F_{tx,s,\theta }}\), \({F_{tx,s,\phi }}\) }& Field patterns of TX element \(s\) along spherical basis vectors \(\hat{\theta}\) and \(\hat{\phi}\), respectively  \\
    \hline
    \multicolumn{2}{c|}{\({\hat{r}_{rx,n,m }}\), \({\hat{r}_{tx,n,m }}\)}& Spherical unit vectors of ray \(m\) in cluster \(n\) at the RX and TX, respectively \\
    \hline
    \multicolumn{2}{c|}{\({\bar{d}_{rx,u}}\) }&  Location vectors of RX element \(u\) and TX element \(s\), respectively\\
    \hline
    \multicolumn{2}{c|}{\({{d}_{\rm{3D}}}\) }& 3D distance between the reference points of the BS and the UE \\
    \hline
    \end{tabular}
    \label{Tab_Par}
    \vspace{-0.1cm}
\end{table*}

\subsection{Empirical Observations and Model Gaps}
\begin{figure}[h!]
\centering
\subfloat[]
{\includegraphics[width=7.5cm]{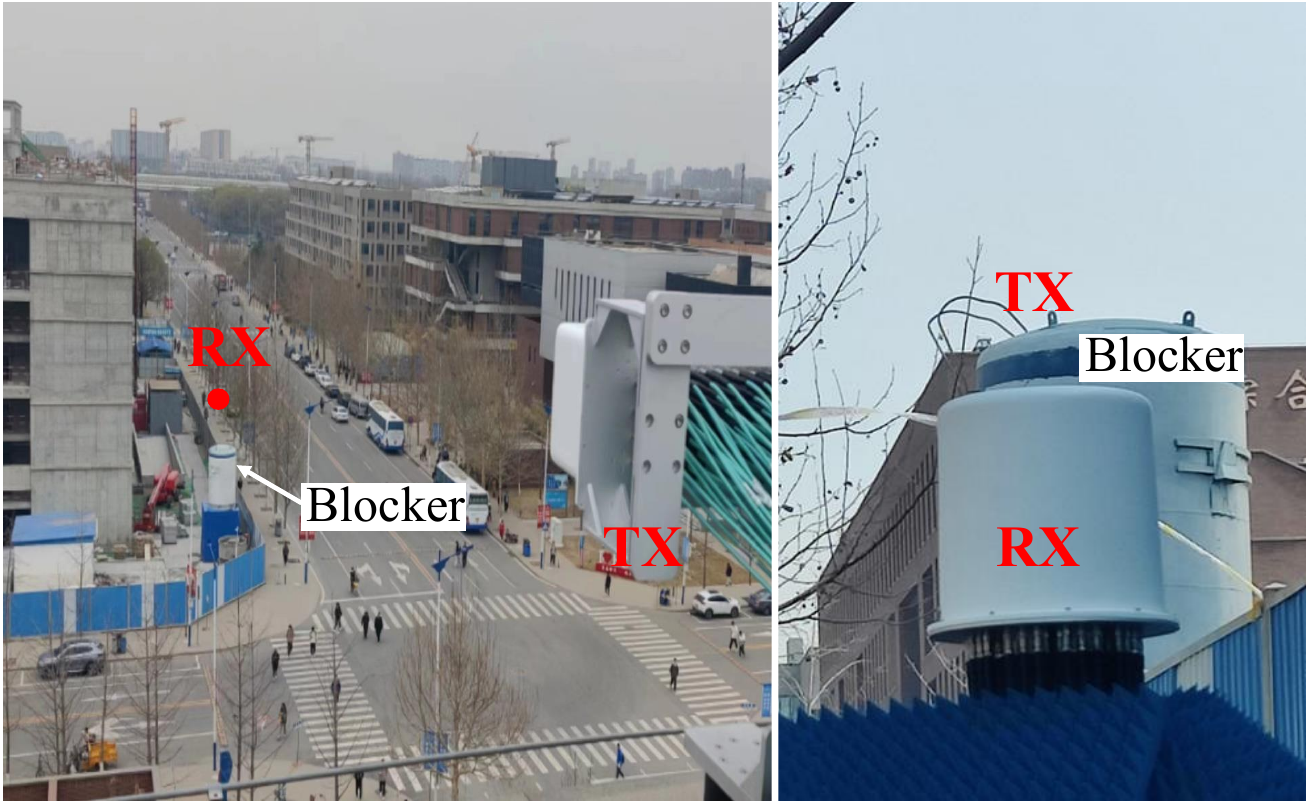}
\label{fig1_a}}
\hfill
\subfloat[]
{\includegraphics[width=7.5cm]{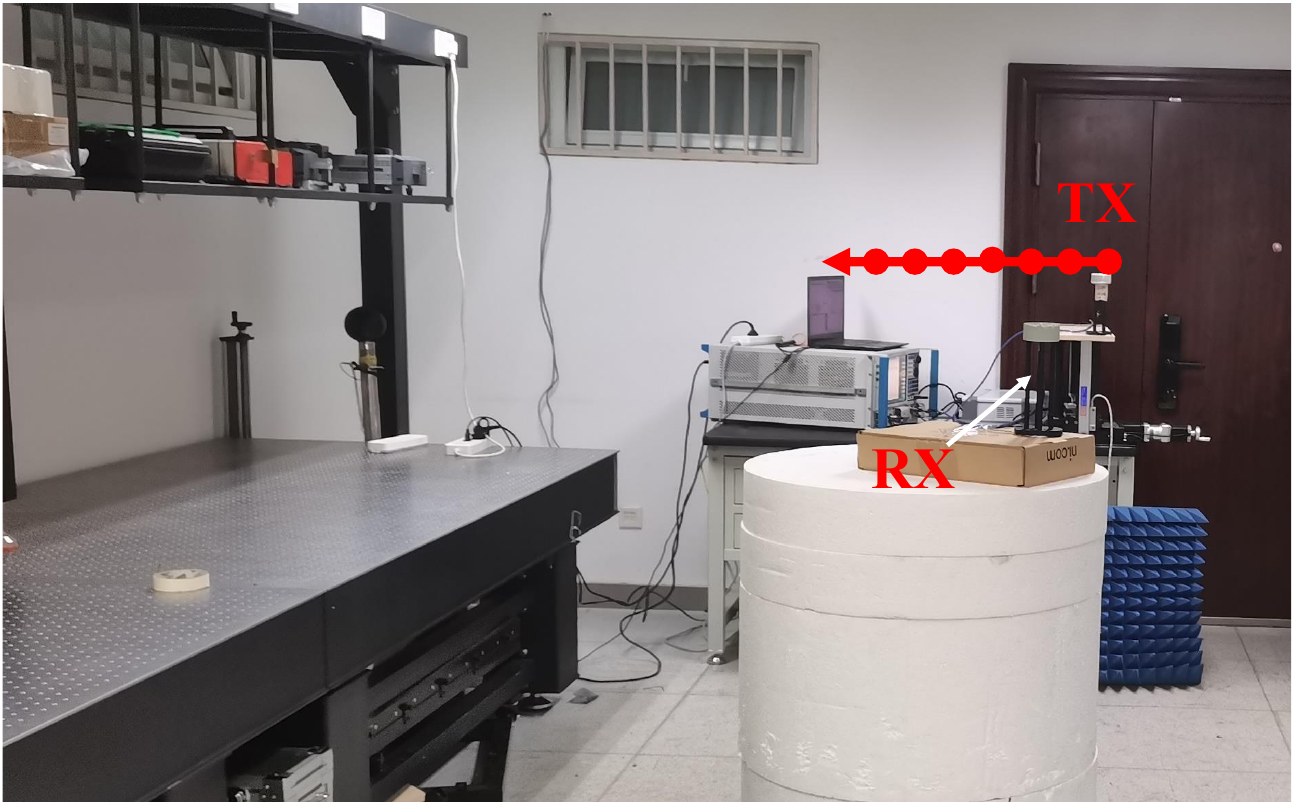}}\label{fig1_b}
\hfill
\caption{Measurement photographs taken in (a) UMa and (b) indoor scenarios.}
\label{fig_mea_scenario}
\end{figure}

To investigate the evolution of channel characteristics from conventional MIMO to XL-MIMO systems, channel measurements were conducted in both urban macro (UMa) and indoor environments, as illustrated in Fig.~\ref{fig_mea_scenario}. In the UMa scenario, a correlation-based time-domain channel sounding platform was used, operating at a center frequency of 13 GHz. The TX was equipped with a uniform planar array (UPA) comprising of $32 \times 2$ dual-polarized antenna elements—i.e., 32 elements in the horizontal and 2 in the vertical dimension. To emulate a larger array, a mechanical sliding platform was utilized to extend the array by four translations horizontally and three vertically, yielding a virtual XL-MIMO array with $128 \times 6$ dual-polarized elements, totaling 1536 antenna elements \cite{3GPP_BUPT_hefei}. In the indoor scenario, where the environment was quasi-static, a frequency-domain channel sounding system based on a vector network analyzer (VNA) was employed. The measurement covered a frequency sweep from 10 to 24 GHz, with a total of 1401 frequency points. The ultra-wide bandwidth of 14 GHz enabled high delay resolution. A virtual uniform linear array (ULA) of 83 elements was synthesized at the TX side by mechanically translating a single omnidirectional antenna in 5 mm increments. The RX employed an identical omnidirectional antenna. Detailed measurement configurations are summarized in Table~\ref{Tab_Mea_Configuration}.

\begin{table}
\renewcommand\arraystretch{1.2}
    \centering
    \caption{Measurement Configurations}
        \begin{tabular}{m{2cm}<{\centering}|m{2.5cm}<{\centering}|m{2.5cm}<{\centering}}
    \hline
       Parameters  & \multicolumn{2}{c}{Values }  \\
          \hline
        Scenario  & UMa & Indoor   \\
         \hline
         Measurement platform  & Correlation-based time-domain platform  & VNA-based frequency-domain platform\\
         \hline
        Center frequency & 13 GHz & 17 GHz   \\
         \hline
        Bandwidth & 400 MHz & 14 GHz   \\
        \hline
        TX array & UPA (128$\times$6, dual-polarization) & ULA (83 elements)  \\
        \hline
        Rx array & Omnidirectional array (64 elements) & Single antenna \\
        \hline
        TX height & 27.8 m & 1.5 m  \\
        \hline
        RX height & 1.8 m & 1.5 m  \\
        \hline
        TX-RX distance & 149.25 m & 0.88 m  \\
        \hline
    \end{tabular}
    \label{Tab_Mea_Configuration}
\end{table}

\begin{figure}[h!]
\centering
\subfloat[]
{\includegraphics[width=8.5cm]{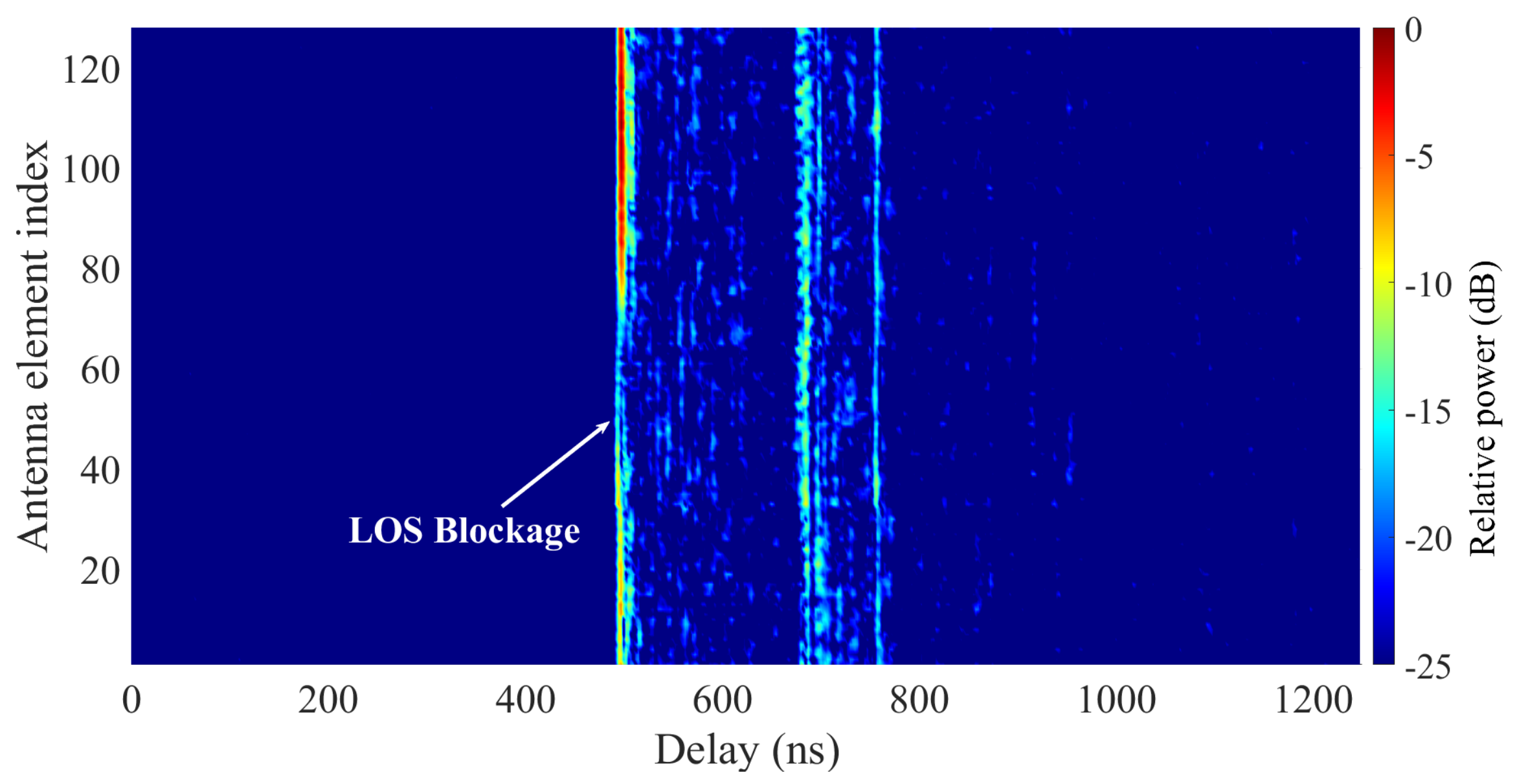}\label{fig_PDP_a}}
\hfill
\subfloat[]
{\includegraphics[width=8.4cm]{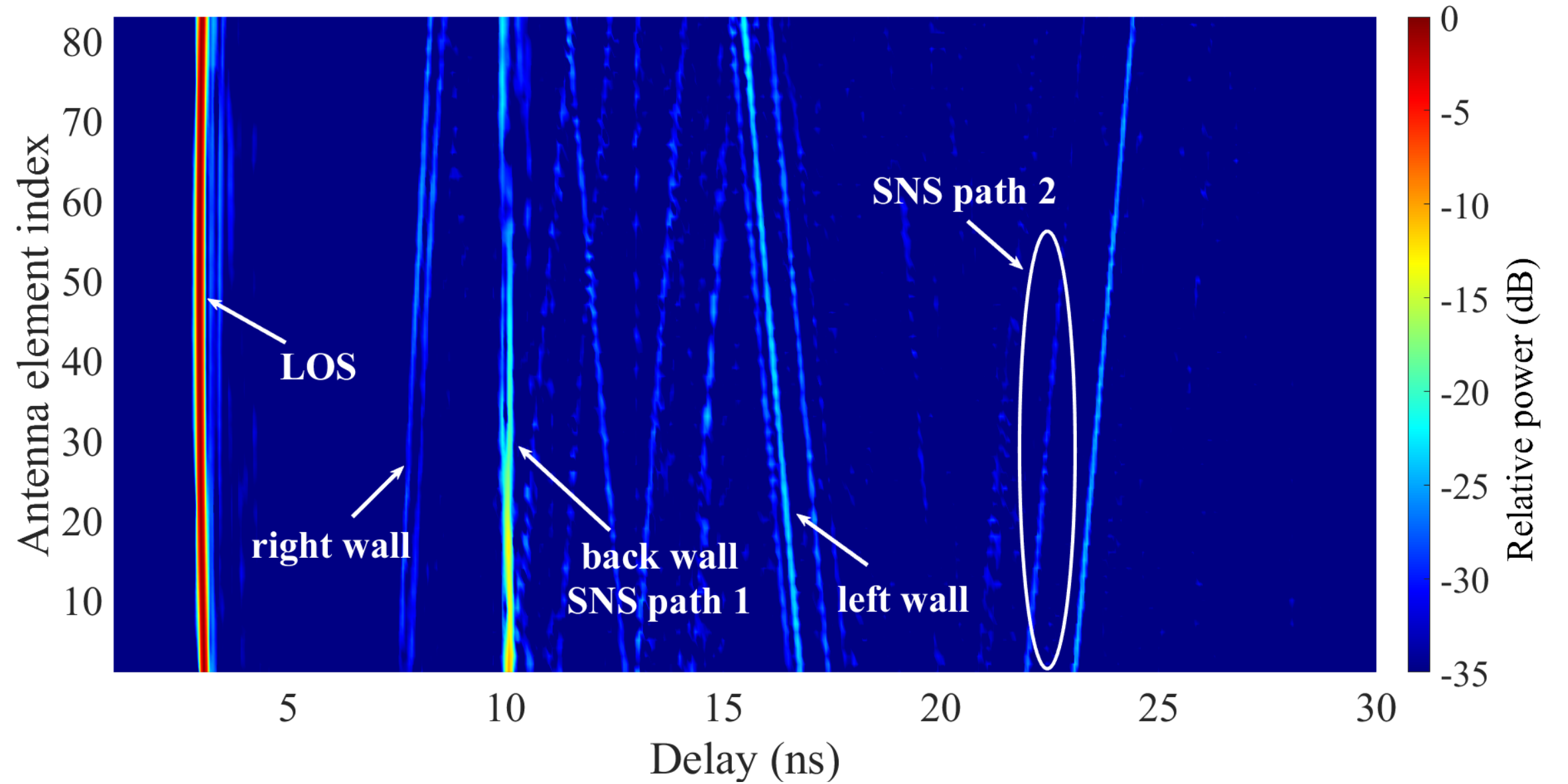}\label{fig_PDP_b}}
\caption{PDPs across TX antenna elements in (a) UMa and (b) indoor scenarios.}
\label{fig_PDP}
\end{figure}

Fig.~\ref{fig_PDP} shows the power delay profiles (PDPs) across the TX elements in both UMa and indoor scenarios, where the horizontal and vertical axes represent the propagation delay and antenna element index, respectively, and the color indicates the relative received power in dB. For the UMa scenario, one row of the virtually synthesized XL-MIMO array is selected, comprising 128 elements. Notably, certain path trajectories exhibit substantial power fluctuations along the array, with some paths even falling below the noise floor. This spatial power variation reveals the presence of SNS. Furthermore, in Fig.~\ref{fig_PDP_b}, sloped path trajectories are observed, reflecting propagation delay differences across antenna elements. This phenomenon arises from the large array aperture and the high delay resolution enabled by the ultra-wide bandwidth, indicating the presence of near-field propagation. A more detailed analysis of near-field propagation and SNS is provided below, based on paths extracted from the measurement data.

\begin{figure}[h!]
\centering
\subfloat[]
{\includegraphics[width=8cm]{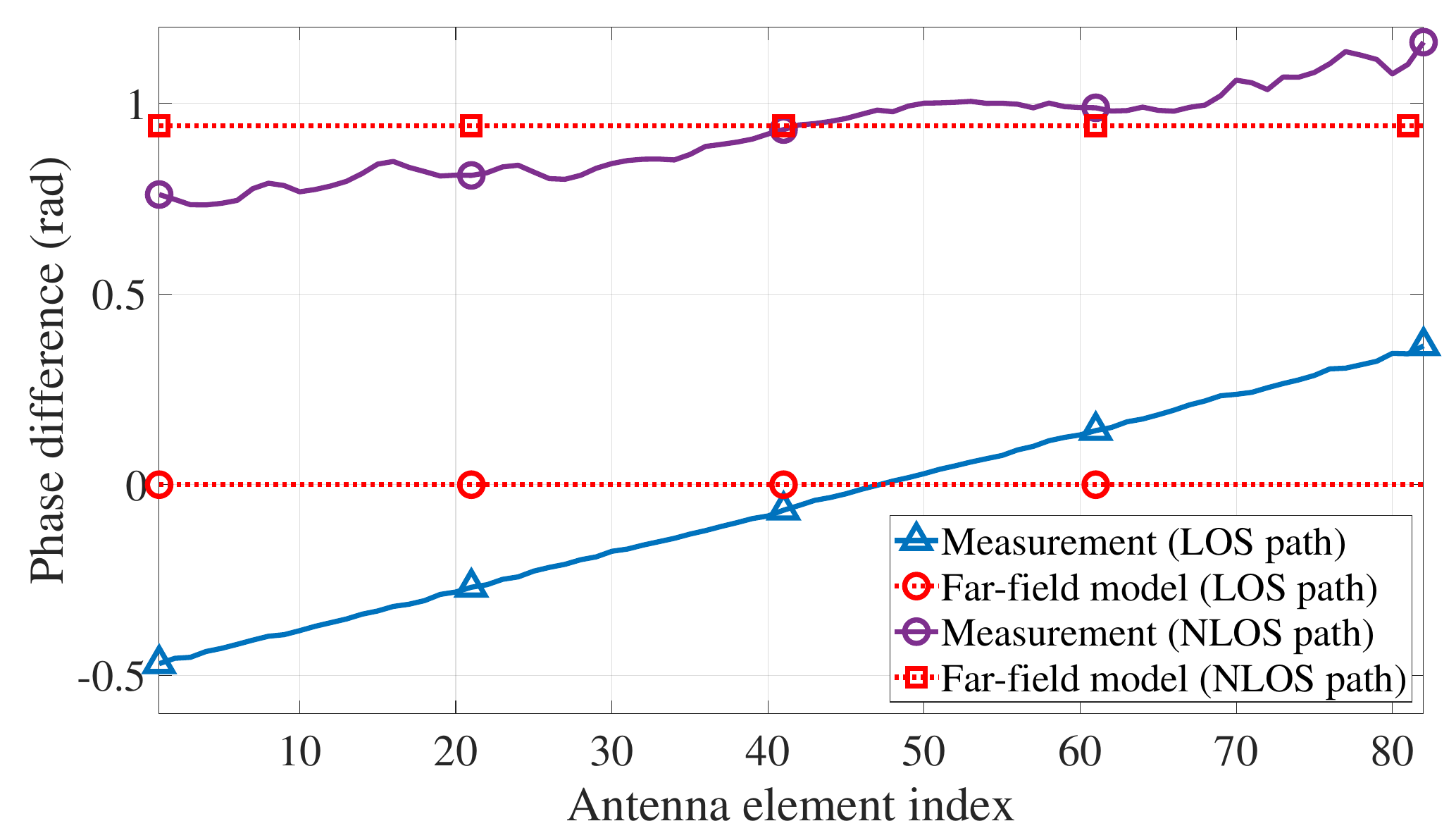} \label{fig_NF_Mea_a}}
\hfill
\subfloat[]
{\includegraphics[width=8cm]{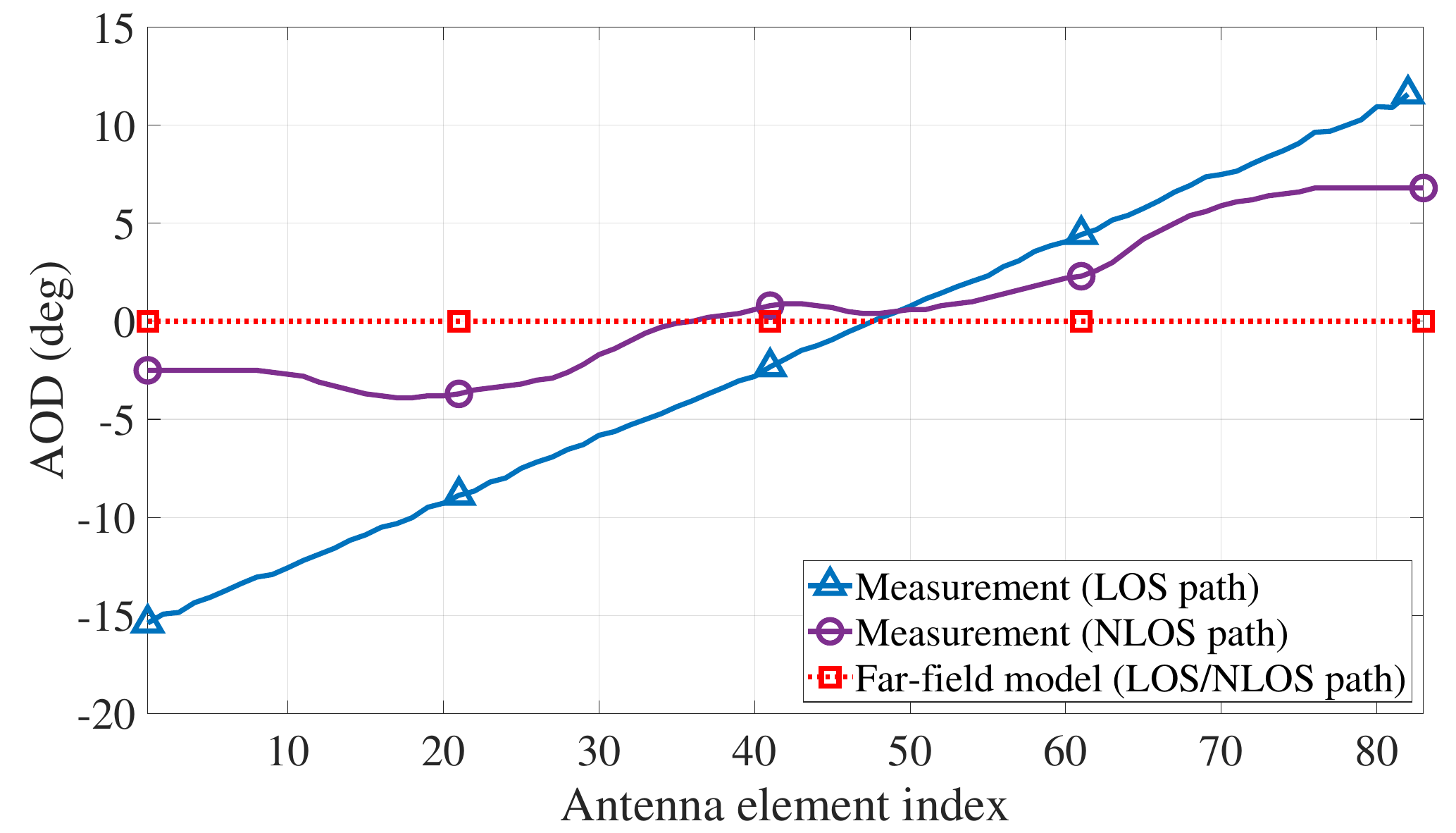} \label{fig_NF_Mea_b}}
\caption{Variations of near-field parameters across TX antenna elements: (a) inter-element phase difference; (b) AODs.}
\label{fig_NF_Mea}
\end{figure}
\textit{Near-Field Propagation}: Given the high delay resolution and superior measurement accuracy of the indoor channel sounding platform—both of which are critical for reliably capturing near-field propagation feature—the indoor measurement data are selected for the analysis. As illustrated in Fig.~\ref{fig_NF_Mea}, the near-field propagation feature manifests as significant variations in both phase and AOD across the antenna elements. Due to the periodic nature of phase, the measured values are wrapped within the interval $[0, 2\pi)$, resulting in a folded appearance that obscures the true spatial variation. To address this, inter-element phase differences are computed and presented in Fig.~\ref{fig_NF_Mea_a}, providing a clearer view of relative phase evolution along the array. Under the far-field assumption, the phase varies linearly with the element index, producing a constant inter-element phase difference. However, measurements exhibit clear deviations from this linearity—particularly for the LOS path, where the inter-element phase difference fluctuates by as much as 0.83 radians. This nonlinearity cannot be captured by far-field models, thereby underscoring the necessity of modeling the near-field feature. In addition, the LOS path exhibits an AOD variation of up to 26.98°, as shown in Fig.~\ref{fig_NF_Mea_b}, which aligns with geometric calculations based on the positions of the TX and RX elements. A comparable level of angular variation is also observed for the NLOS path reflected from the back wall. Similar angular behavior has been reported in \cite{3GPP_ZTE_hefei} as well. These spatially varying angular parameters, typically omitted in far-field models that assume uniform AODs across the array, need to be incorporated into modeling frameworks to ensure realism.

\begin{figure}[htbp]
\centering
\subfloat[]
{\includegraphics[width=8cm]{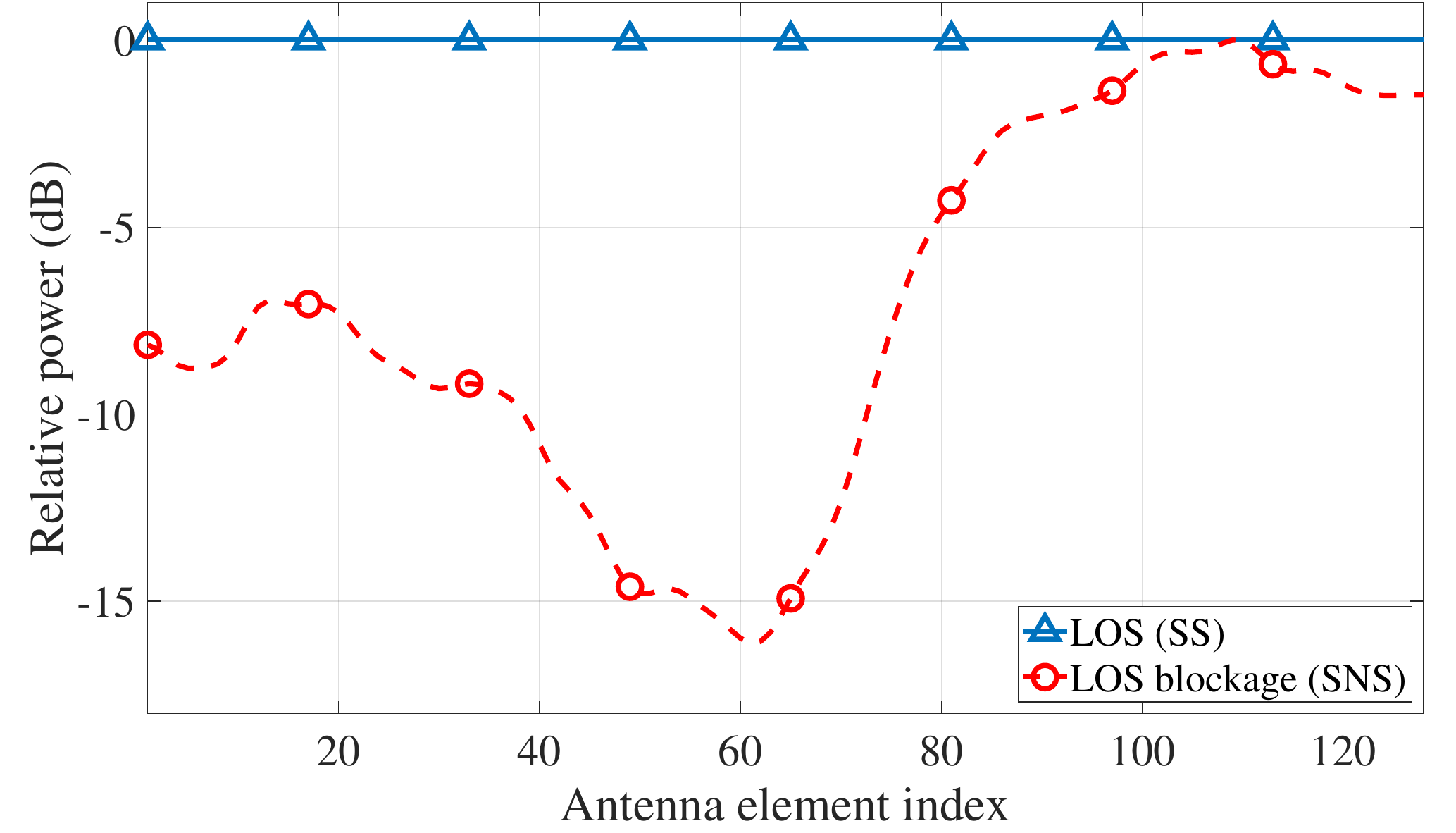} \label{fig_SNS_Mea_a} }
\hfill
\subfloat[]
{\includegraphics[width=8cm]{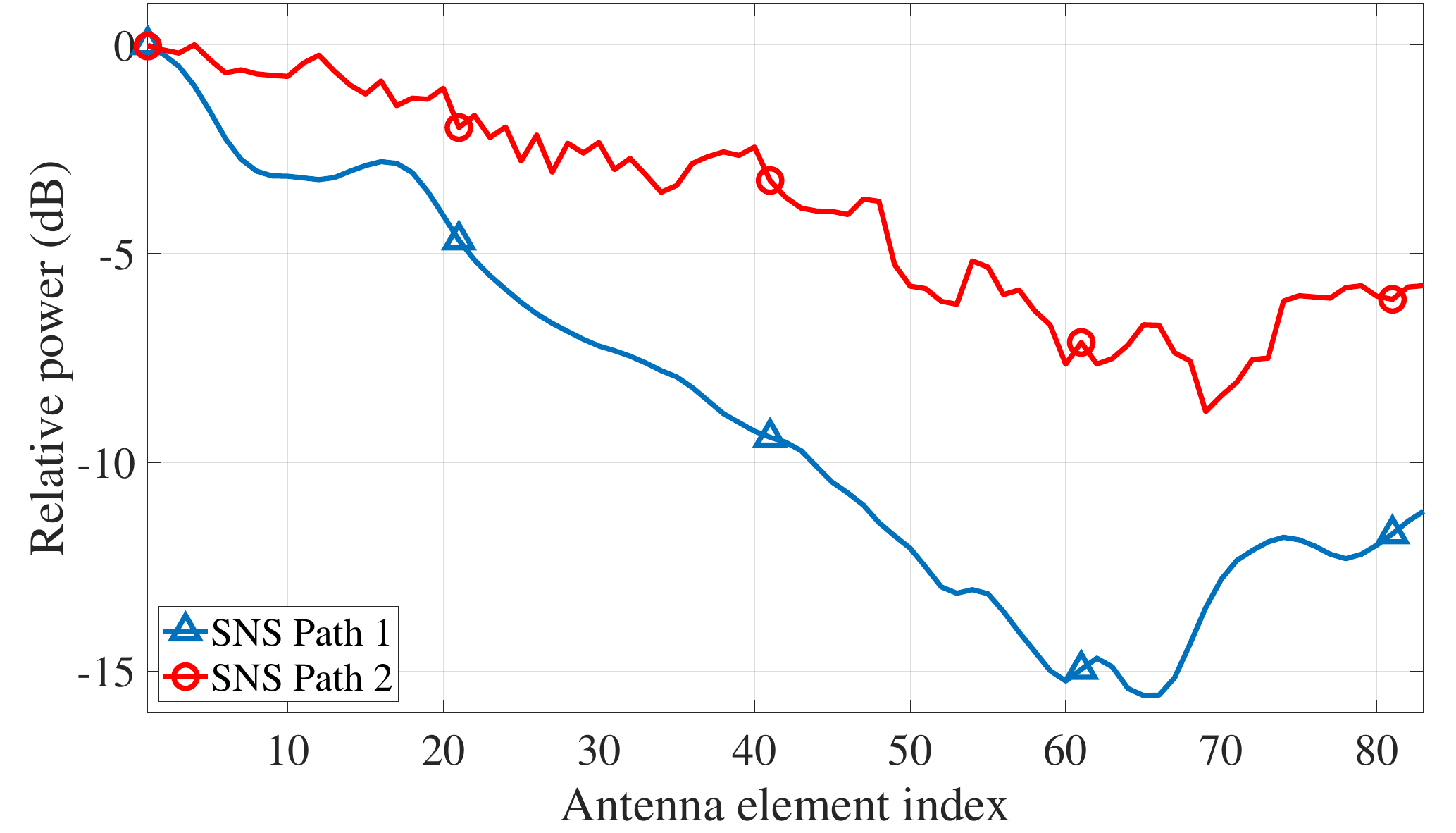} \label{fig_SNS_Mea_b} }
\caption{Power variation of SNS paths across TX antenna elements in (a) UMa and (b) indoor scenarios.}
\label{fig_SNS_Mea}
\end{figure}

\textit{Spatial Non-Stationarity:} In the UMa scenario, a power fluctuation of up to 16.13 dB was observed in the LOS path under partial blockage, indicating that such obstructions are a key contributor to SNS. Similar element-wise power variations due to partial blockage by building edges were also reported in \cite{3GPP_Eri_helan}. In addition, some paths exhibit notable element-wise power variations, even in the absence of blockage, as shown in Fig.~\ref{fig_SNS_Mea_b}. This behavior results from incomplete scattering, where scatterers with limited physical size fail to uniformly illuminate the entire array. As a result, certain paths may appear strong over some elements but become weak or even undetectable over others. These observations highlight the importance of incorporating the SNS feature into channel models for XL-MIMO systems, in order to accurately capture the element-wise power variations of SNS paths.

\subsection{XL-MIMO Channel Modeling Framework}
\begin{figure*}[htbp]
\centering
\includegraphics[width=16 cm]{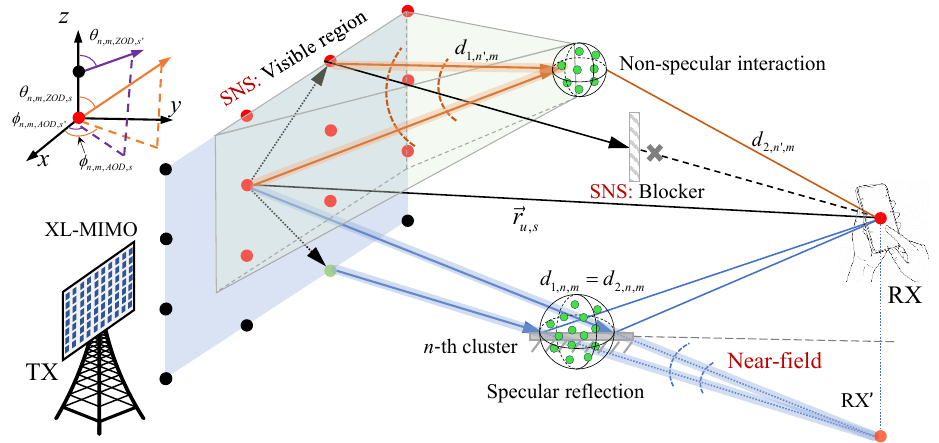}
\caption{Illustration of the XL-MIMO channel modeling framework incorporating the near-field propagation and SNS features.}
\label{fig_general}
\end{figure*}

A schematic of the XL-MIMO channel modeling framework, which jointly incorporates near-field propagation and SNS features, is shown in Fig.~\ref{fig_general}. Accordingly, the LOS and NLOS channel coefficients in the MIMO channel model are modified as given in Eqs.~(\ref{equ_H_LOS_6G}) and~(\ref{equ_H_NLOS_6G}), respectively. All variables associated with near-field propagation—including element-wise angular parameters $\phi_{\mathrm{LOS}, \mathrm{AOD}, u,s}$, $\phi_{\mathrm{LOS}, \mathrm{AOA}, u,s}$, $\theta_{\mathrm{LOS}, \mathrm{ZOD}, u,s}$, $\theta_{\mathrm{LOS}, \mathrm{ZOA}, u,s}$, $\phi_{n,m,\mathrm{AOD},s}$, $\phi_{n,m,\mathrm{AOA},u}$, $\theta_{n,m,\mathrm{ZOD},s}$, $\theta_{n,m,\mathrm{ZOA},u}$, the geometric vector $\vec{r}_{u,s}$, and distances $d_{1,n,m}$ and $d_{2,n,m}$—as well as the SNS-related parameters $\alpha_{s,n,m}^{\mathrm{SNS}}$ and $\beta_u^{\mathrm{SNS}}$, are explicitly listed in Table~\ref{Tab_Par}. For clarity, all parameters in this paper are defined under the downlink assumption, where the BS serves as the TX and the UE acts as the RX.

\begin{figure*}
\begin{equation}
\begin{aligned}
H_{u,s,1}^{\rm{LOS}}&\left( t \right) =  \sqrt{\beta_{u}^{\rm{SNS}} \alpha_{s,1}^{\rm{SNS}}} \left[ {\begin{array}{*{20}{c}}
{{F_{rx,u,\theta }}\left( {{\theta _{{\rm{LOS,ZOA}},u,s}},{\phi _{{\rm{LOS,AOA}},u,s}}} \right)}\\
{{F_{rx,u,\phi }}\left( {{\theta _{{\rm{LOS,ZOA}},u,s}},{\phi _{{\rm{LOS,AOA}},u,s}}} \right)}
\end{array}} \right]\left[ {\begin{array}{*{20}{c}}
1&0\\
0&{ - 1}
\end{array}} \right] \\ 
&\cdot \left[ {\begin{array}{*{20}{c}}
{{F_{tx,s,\theta }}\left( {{\theta _{{\rm{LOS,ZOD}},u,s}},{\phi _{{\rm{LOS,AOD}},u,s}}} \right)}\\
{{F_{tx,s,\phi }}\left( {{\theta _{{\rm{LOS,ZOD}},u,s}},{\phi _{{\rm{LOS,AOD}},u,s}}} \right)}
\end{array}} \right]  \exp \left( { - j2\pi \frac{{{d_{{\rm{3D}}}}}}{{{\lambda _0}}}} \right)\exp \left( { - j2\pi \frac{{\left| {{{\vec r}_{u,s}}} \right| - {d_{{\rm{3D}}}}}}{{{\lambda _0}}}} \right)\exp \left( {j2\pi \frac{{\hat r_{rx,{\rm{LOS}}}^T\bar v}}{{{\lambda _0}}}t} \right).
\end{aligned} 
\label{equ_H_LOS_6G}
\end{equation}
\end{figure*}

\begin{figure*}
\begin{equation}
\begin{aligned}
H&_{u,s,n,m}^{\rm{NLOS}}\left( t \right)  = \sqrt{\beta_{u}^{\rm{SNS}} \alpha_{s,n,m}^{\rm{SNS}}} \sqrt {\frac{{{P_n}}}{M}} \left[ {\begin{array}{*{20}{c}}
{{F_{rx,u,\theta }}\left( {{\theta _{n,m,{\rm{ZOA}},u}},{\phi _{n,m,{\rm{AOA}},u}}} \right)}\\
{{F_{rx,u,\phi }}\left( {{\theta _{n,m,{\rm{ZOA}},u}},{\phi _{n,m,{\rm{AOA}},u}}} \right)}
\end{array}} \right]\mathbf{\Phi}_{n,m} \left[ {\begin{array}{*{20}{c}}
{{F_{tx,s,\theta }}\left( {{\theta _{n,m,{\rm{ZOD}},s}},{\phi _{n,m,{\rm{AOD}},s}}} \right)}\\
{{F_{tx,s,\phi }}\left( {{\theta _{n,m,{\rm{ZOD}},s}},{\phi _{n,m,{\rm{AOD}},s}}} \right)}
\end{array}} \right] \\ 
& \cdot  \exp \left( {j2\pi \frac{{{d_{2,n,m}} - \left\| {{d_{2,n,m}} \cdot {{\hat r}_{rx,n,m}} - {{\bar d}_{rx,u}}} \right\|}}{{{\lambda _0}}}} \right)\exp \left( {j2\pi \frac{{{d_{1,n,m}} - \left\| {{d_{1,n,m}} \cdot {{\hat r}_{tx,n,m}} - {{\bar d}_{tx,s}}} \right\|}}{{{\lambda _0}}}} \right)\exp \left( {j2\pi \frac{{\hat r_{rx,n,m}^T\bar v}}{{{\lambda _0}}}t} \right).
\end{aligned} 
\label{equ_H_NLOS_6G}
\end{equation}
\end{figure*}

\subsubsection{Near-field Propagation}

Compared to the far-field model, near-field propagation introduces two key improvements: the array phase and angular-domain parameters, both for the direct path and non-direct paths\footnote{The direct path refers to the LOS ray as defined in 3GPP TR 38.901 \cite{C3_3GPP}, while non-direct paths include all remaining clusters/rays excluding the LOS ray.}. For the direct path, the conventional far-field model approximates the array phases using LOS direction unit vectors $\hat{r}_{\mathrm{tx, LOS}}$ and $\hat{r}_{\mathrm{rx, LOS}}$. In contrast, the near-field model calculates the exact propagation distance \(\left| {{{\vec r}_{u,s}}} \right|\) between each antenna element pair $(s, u)$, enabling more accurate phase modeling. Moreover, angular-domain parameters, which are typically assumed constant across the array in far-field models, are now treated as element-wise variables that depend on the spatial positions of the TX and RX antenna elements. For the non-direct paths, two additional parameters—$d_{1,n,m}$ and $d_{2,n,m}$—are introduced to represent the distances from the BS and UE, respectively, to a spherical-wave source of the $m$-th ray in cluster $n$. These distances are used to compute the element-wise phase terms. Similar to the direct path, the angular-domain parameters for non-direct paths are modeled as element-wise variables. Specifically, the near-field angular parameters are parameterized by the indices of the TX element $s$ and RX element $u$.

Other parameters—namely amplitude, XPR, delay, and Doppler shift—are retained from the far-field model without modification. Prior studies have demonstrated that significant amplitude variations across array elements generally occur only when the TX–RX separation is comparable to the array aperture \cite{C2_Bjorson_dis,C2_UPD}. However, this condition is rarely satisfied in practical scenarios, where the communication distance is typically much larger than the array size. Therefore, amplitude differences under near-field propagation can generally be considered negligible. Similarly, XPR variation across antenna pairs is commonly omitted. Delay variations across elements are often ignored in channel modeling, as they are typically much smaller than the system sampling interval and are inherently coupled with phase, making separate modeling unnecessary \cite{3GPP_vivo_helan}. As for Doppler shift, most near-field scenarios involve low-mobility handheld UEs, resulting in small Doppler spreads; hence, element-wise Doppler shift modeling is generally unnecessary \cite{3GPP_QC_US}.

\subsubsection{Spatial Non-Stationarity}
Compared to the SS model, the SNS model captures path power variation across antenna elements at either the BS or UE side by introducing element-wise power attenuation factors. At the BS side, $\alpha_{s,n,m}^{\mathrm{SNS}}$ represents the power variation of the $m$-th ray in the $n$-th cluster across the TX antenna elements. At the UE side, power variation among antenna elements may result from hand blockage or proximity to the user's head during voice communication. Since such obstructions are typically close to the UE antenna array, their impact tends to be uniform across all clusters. To reflect this effect, a UE-side attenuation factor, $\beta_{u}^{\mathrm{SNS}}$, is introduced and applied identically across all clusters for each UE antenna element.

\section{Near-field Propagation Feature}

\begin{figure*}[b]
    \begin{equation}
        H_{u,s}^{\rm{NLOS}}\left( {\tau ,t} \right) = \underbrace {\underbrace {\sum\limits_{n = 1}^2 {\sum\limits_{i = 1}^3 {\sum\limits_{m \in {R_i}} {H_{u,s,n,m}^{\rm{NLOS}}(t)\delta \left( {\tau  - {\tau _{n,i}}} \right) + } } } }_{{\rm{two \ strongest \ clusters}}}\sum\limits_{n = 3}^{{N_{\rm{spec}}}} {\sum\limits_{m = 1}^M {H_{u,s,n,m}^{\rm{NLOS}}(t)\delta \left( {\tau  - {\tau _n}} \right)} } }_{{\rm{specular \ reflection \ clusters}}} + \underbrace {\sum\limits_{n = {N_{\rm{spec}}} + 1}^{N - {N_{spec}}} {\sum\limits_{m = 1}^M {H_{u,s,n,m}^{\rm{NLOS}}(t)\delta \left( {\tau  - {\tau _n}} \right)} } }_{{\rm{non - specular \ reflection \ clusters}}}.
        \label{equ_NF_NLOS}
    \end{equation}
\end{figure*}

Within the XL-MIMO channel modeling framework, we now further discuss the generation of near-field propagation parameters. For the direct path, since the antenna element locations at both the BS and UE are explicitly known in MIMO channel models, the element-wise parameters can be obtained directly via geometric calculations. Therefore, detailed derivation for the direct path is omitted here.

\begin{figure*}[h]
\centering
\subfloat[]
{\includegraphics[width=8cm]{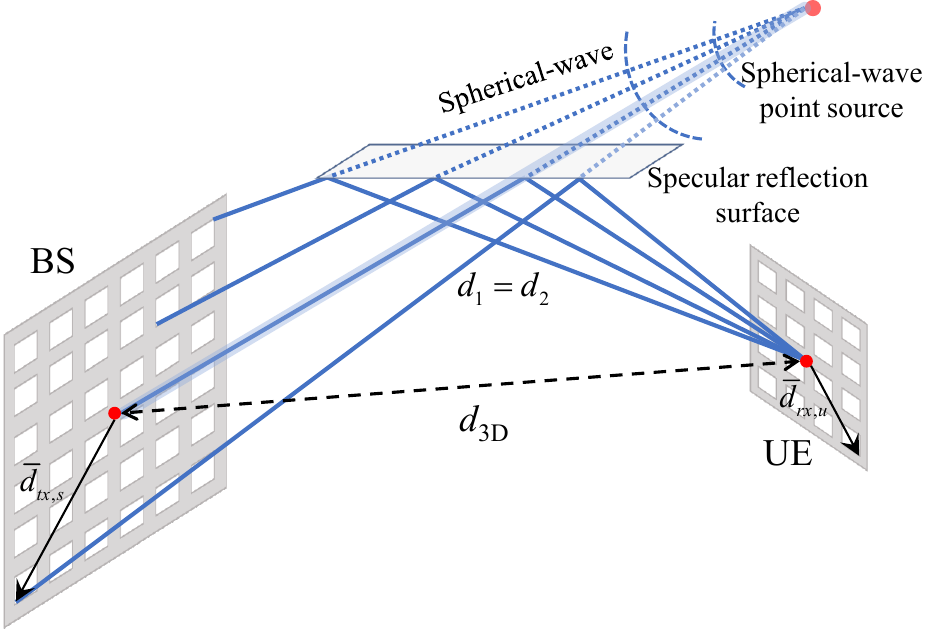}\label{fig_NF_model_a}}
\subfloat[]
{\includegraphics[width=7.5cm]{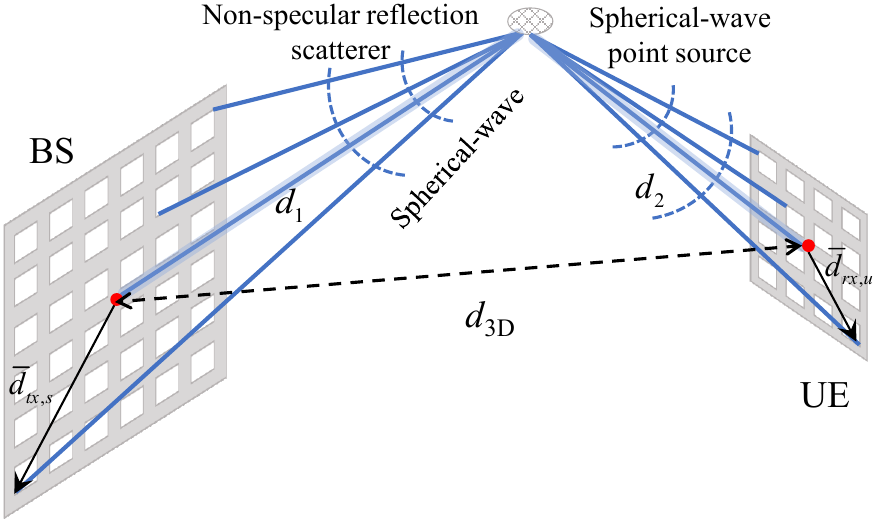} \label{fig_NF_model_b}}
\caption{Near-field propagation feature: (a) Specular reflection and (b) Non-specular reflection.}
\vspace{-0.5cm}
\label{fig_NF_model}
\end{figure*}

For non-direct paths, it is necessary to determine the distances from the BS and UE to the spherical-wave source associated with each cluster, denoted as $d_1$ and $d_2$, respectively. It is important to note that $d_1$ and $d_2$ differ from the scatterer-to-BS/UE distances commonly assumed in conventional models, as scatterers may not behave as ideal point sources of spherical waves.

To address this, the near-field NLOS propagation model is formulated as shown in Eq.~(\ref{equ_NF_NLOS}), where all clusters are categorized into specular and non-specular reflections. The specular reflection clusters are defined as the $N_{\mathrm{spec}}$ strongest clusters, i.e., those indexed by $n = 1, 2, \ldots, N_{\mathrm{spec}}$. The locations of the spherical-wave sources for both cluster types are depicted in Fig.~\ref{fig_NF_model}. For specular reflection clusters, the spherical-wave source at the BS side is the mirror image of the UE reference point with respect to the reflecting surface. Conversely, at the UE side, the spherical-wave source is the mirror image of the BS reference point. Fig.~\ref{fig_NF_model} illustrates only the BS-side case. Therefore, $d_1$ and $d_2$ correspond to the total propagation path length, calculated as the absolute delay multiplied by the speed of light. Assuming the scatterer itself as the wave source would underestimate these distances. In contrast, for non-specular reflection clusters, the spherical-wave source is assumed to be located at the actual physical position of the scatterer. The following subsections detail the generation key parameters: the number of specular clusters, and the distances from the spherical-wave sources to the BS and UE.

\subsection{Number of Specular Reflection Clusters}

To determine the number of specular reflection clusters, we first define a scaling factor at the BS side as ${s_{\mathrm{BS}}} = d_1 / (c \tilde{\tau})$, where $c$ is the speed of light, and $\tilde{\tau}$ is the absolute propagation delay of the path. The scaling factor ${s_{\mathrm{BS}}}$ takes values between 0 and 1, where ${s_{\mathrm{BS}}} = 1$ corresponds to LOS or purely specular reflection paths, and ${s_{\mathrm{BS}}} < 1$ indicates the presence of non-specular mechanisms, such as diffraction or diffuse scattering. This factor reflects the relative position of the non-specular interaction point—smaller values indicate that the interaction occurs closer to the BS, while values approaching 1 suggest proximity to the UE.

\begin{figure}[htbp]
\centering
\subfloat[]{
{\includegraphics[width=4.3cm]{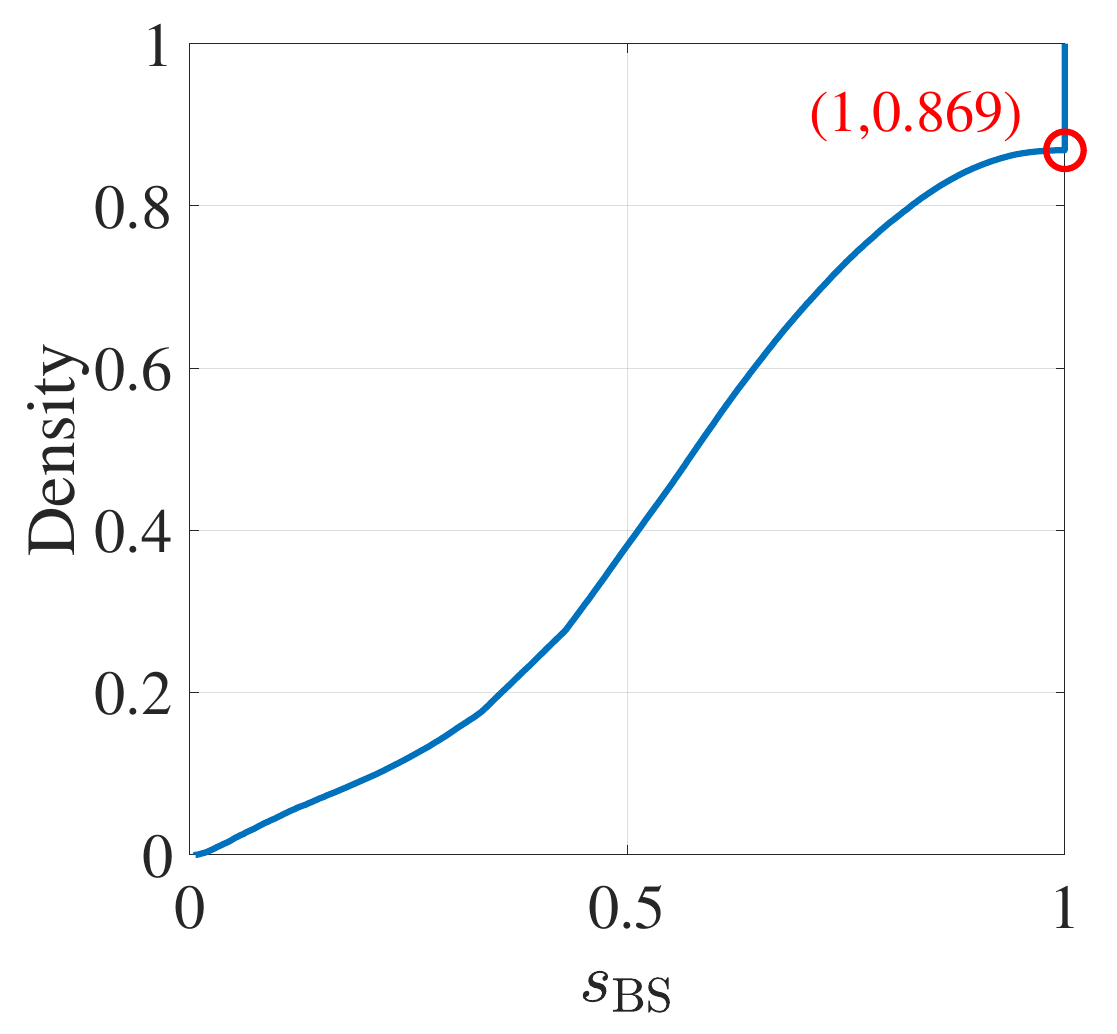}}}
\subfloat[]{
{\includegraphics[width=4.3cm]{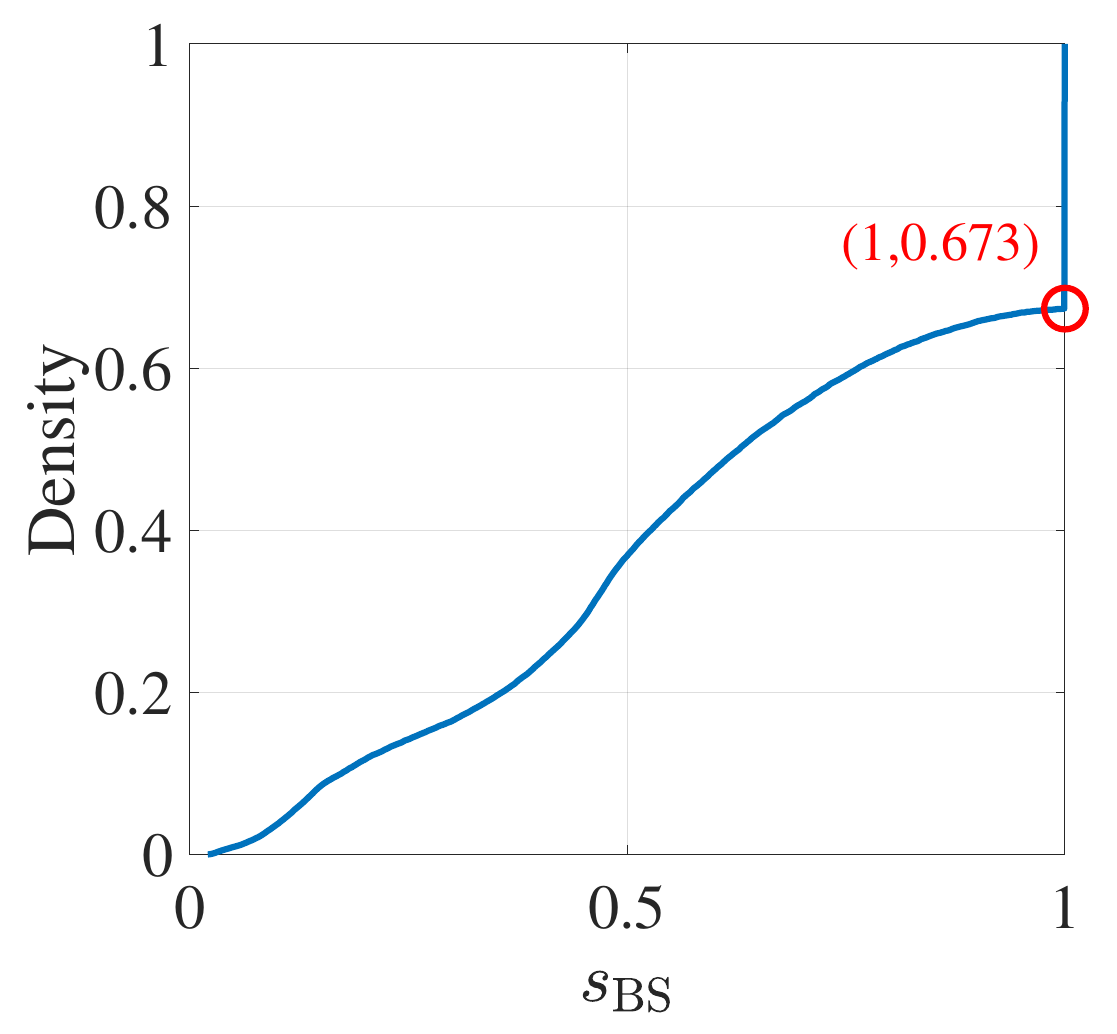}}
}
\caption{CDFs of the scaling factors at the BS side in (a) UMi and (b) InH scenarios.}
\vspace{-0.5cm}
\label{fig_NF_s_TRP}
\end{figure}

Fig.~\ref{fig_NF_s_TRP} shows the cumulative distribution function (CDF) of ${s_{\mathrm{BS}}}$, derived from ray-tracing simulations in typical urban micro (UMi) and indoor hotspot (InH) scenarios, excluding LOS paths. In the UMi scenario, approximately 13.1\% of the paths have ${s_{\mathrm{BS}}} = 1$, indicating specular reflections. In contrast, the InH scenario shows a higher proportion—up to 32.7\%—due to abundant flat surfaces such as walls, floors, and ceilings. Based on these ratios and the total number of clusters specified in the 3GPP channel model \cite{C3_3GPP}, the number of specular clusters is set to two for UMi and six for InH. At the 3GPP RAN1 \#120 bis meeting, there was a consensus on setting the number of specular clusters to two for the UMi scenario. However, most companies preferred reducing the number of specular clusters in the InH scenario to four, despite simulation results from some companies showing that up to 75\% of the paths were specular \cite{3GPP_Eri_wuhan}. For consistency, the UMa and indoor factory (InF) scenarios adopt the same number of specular clusters as UMi and InH, respectively.

It can be inferred that paths consisting purely of specular reflections tend to exhibit higher average power than those involving diffraction or diffuse scattering. This observation is supported by the findings in \cite{3GPP_Eri_wuhan}, where the received power of paths with ${s_{\mathrm{BS}}} = 1$ was found to be stronger than that of paths with ${s_{\mathrm{BS}}} < 1$. Consequently, a practical modeling approach is to designate the $N_{\mathrm{spec}}$ strongest-power clusters as specular reflection clusters.

\begin{table}
\renewcommand\arraystretch{1.2}
    \centering
    \caption{Parameters of the near-field propagation feature}
    \begin{tabular}{m{1.5cm}<{\centering}|m{0.5cm}<{\centering}|m{1cm}<{\centering}|m{1cm}<{\centering}|m{1cm}<{\centering}|m{1cm}<{\centering}}
    \hline
       \multicolumn{2}{c|}{Scenario}  & UMa & UMi & InH & InF\\
         \hline
        \multicolumn{2}{c|}{\(N_{\rm{spec}}\)} & 2 & 2 & 4 & 4  \\
        \hline
        \multirow{2}{*}{\makecell{Beta \\ distribution}} & \(\alpha\) & 1.93 &1.53 & 1.25 & 1.38   \\
        \cline{2-6}
       & \(\beta\) & 1.33 &1.42 & 1.27 & 1.26  \\
       \hline
    \end{tabular}
    \label{Tab_NF}
\end{table}

\subsection{Distances from Spherical-Wave Sources to BS and UE}

The distance from the spherical-wave source to the BS is calculated by multiplying the scaling factor $s_{\mathrm{BS}}$ with the total path length, i.e., $d_1 = s_{\mathrm{BS}} \cdot c \tilde{\tau}$, and is given by:
\begin{equation}
    {d_{1,n,m}} = \left\{ {\begin{array}{*{20}{c}}
{{s_{{\rm{BS}},n}}\left( {{d_{{\rm{3D}}}} + {\tau _n} \cdot c + \Delta \tau  \cdot c} \right),}&{n = 1,2},\\
{{s_{{\rm{BS}},n}}\left( {{d_{{\rm{3D}}}} + {\tau _{n,i}} \cdot c + \Delta \tau  \cdot c} \right),}&{{\rm{others}}},
\end{array}} \right.
\label{equ_NF_d1}
\end{equation}
where $\Delta \tau$ is the excess delay in NLOS condition, it is generated from a lognormal distribution using parameters specified in Table~7.6.9-1 of 3GPP TR 38.901 \cite{C3_3GPP}. The delay \(\tau_{n,i}\) corresponds to the $i$-th sub-cluster delay. 

\begin{figure}[htbp]
\centering
\subfloat[]{
{\includegraphics[width=4.3cm]{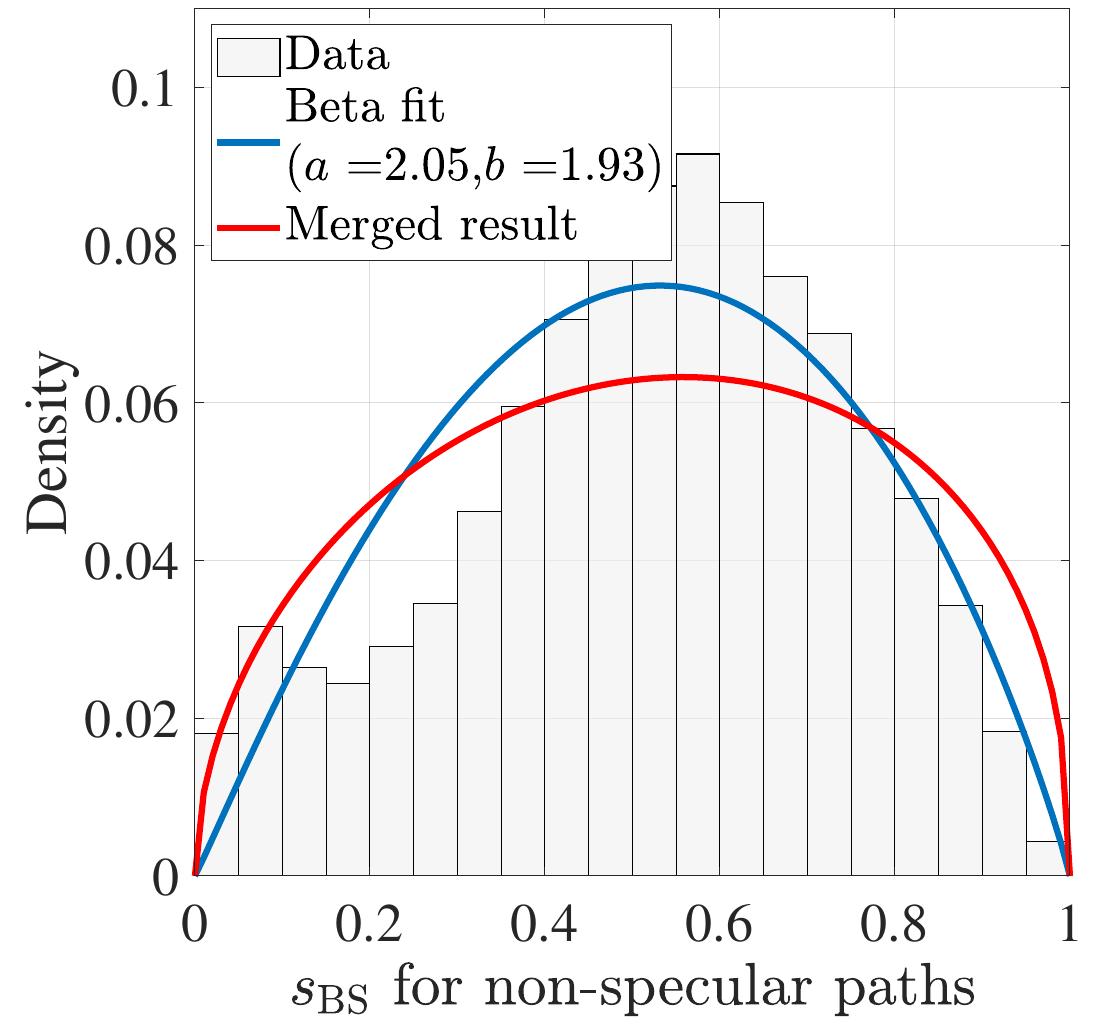}}}
\subfloat[]{
{\includegraphics[width=4.3cm]{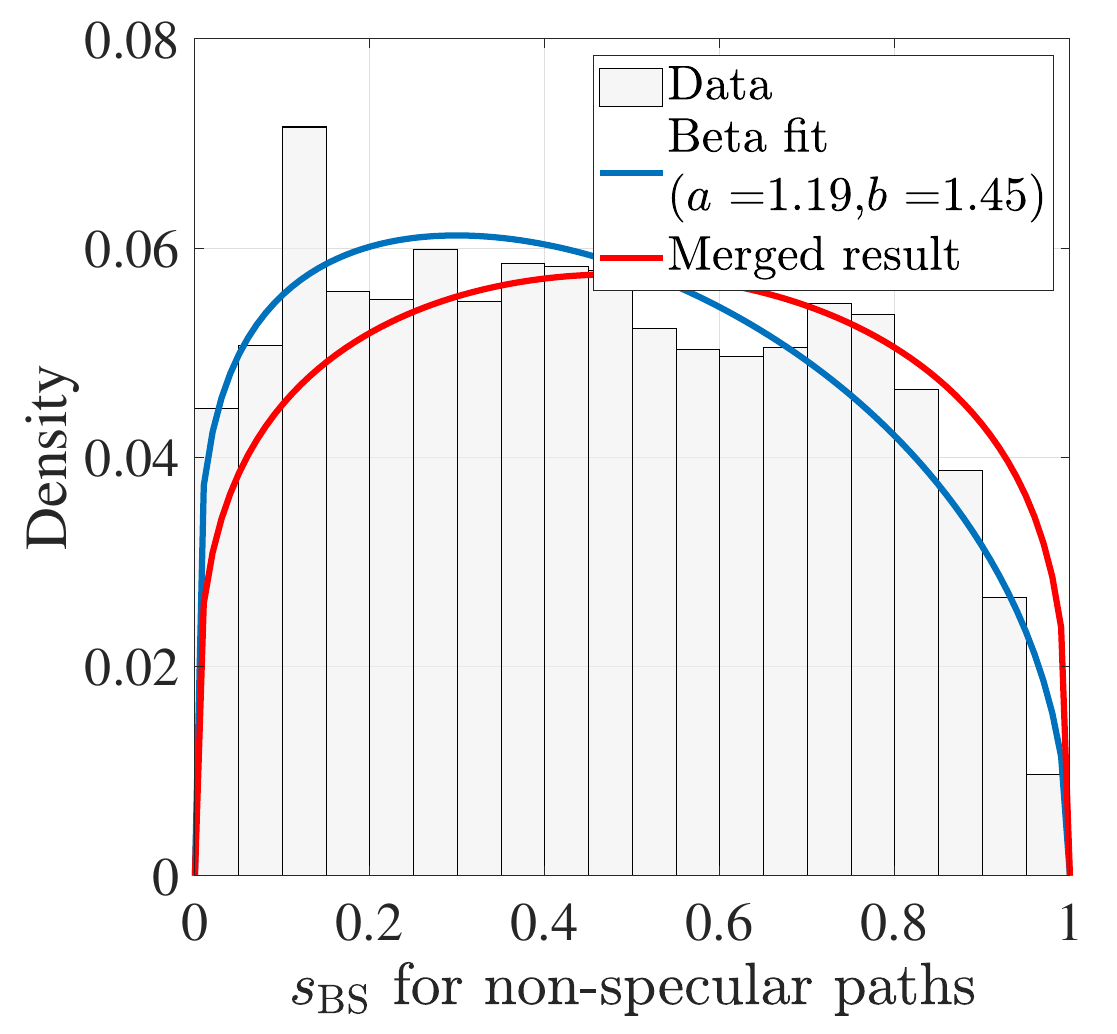}}
}
\caption{Histograms of the scaling factors for non-specular reflection clusters in (a) UMi and (b) InH scenarios.}
\vspace{-0.3cm}
\label{fig_NF_s_TRP_NLOS}
\end{figure}

As established in the previous subsection, if the $n$-th cluster is among the $N_{\mathrm{spec}}$ strongest clusters, the scaling factor is set to $s_{\mathrm{BS}} = 1$. For other clusters, we investigate the stochastic distribution of $s_{\mathrm{BS}}$ excluding values equal to 1. Fig.~\ref{fig_NF_s_TRP_NLOS} presents the histogram of scaling factors in UMi and InH scenarios. The distributions can be well approximated by Beta distributions. Therefore, the scaling factor $s_{{\rm{BS}},n}$ for the \(n\)-th cluster can be expressed as:
\begin{equation}
    {s_{{\rm{BS}},n}} = \left\{ {\begin{array}{*{20}{c}}
1&{n \le N_{\rm{spec}}},\\
{\rm{Beta}(\alpha,\beta)}&{n > N_{\rm{spec}}}.
\end{array}} \right.
\end{equation}

In the UMi scenario, the distribution is left-skewed, indicating that non-specular interactions tend to occur closer to the UE, consistent with the observation that urban scatterers are often located near the UE. In contrast, the InH distribution is more uniform, reflecting a typical indoor environment where both BS and UE are surrounded by scatterers. Merged results from multiple companies also exhibit similar distribution trends \cite{3GPP_summary_wuhan}. Parameters for other scenarios are listed in Table~\ref{Tab_NF}.

Although the aperture size at the UE side is generally smaller than that at the BS side, nearby scatterers in dense environments can still induce near-field effects. To model this, we introduce a scaling factor at the UE side, denoted by $s_{\mathrm{UE}}$, and calculate the corresponding spherical-wave source distance as follows:
\begin{equation}
    {d_{2,n,m}} = \left\{ {\begin{array}{*{20}{c}}
{{s_{{\rm{UE}},n}}\left( {{d_{{\rm{3D}}}} + {\tau _n} \cdot c + \Delta \tau  \cdot c} \right),}&{n = 1,2},\\
{{s_{{\rm{UE}},n}}\left( {{d_{{\rm{3D}}}} + {\tau _{n,i}} \cdot c + \Delta \tau  \cdot c} \right),}&{{\rm{others}}}.
\end{array}} \right.
\label{equ_NF_d1}
\end{equation}

\begin{figure}[htbp]
\centering
\subfloat[]{
{\includegraphics[width=4.3cm]{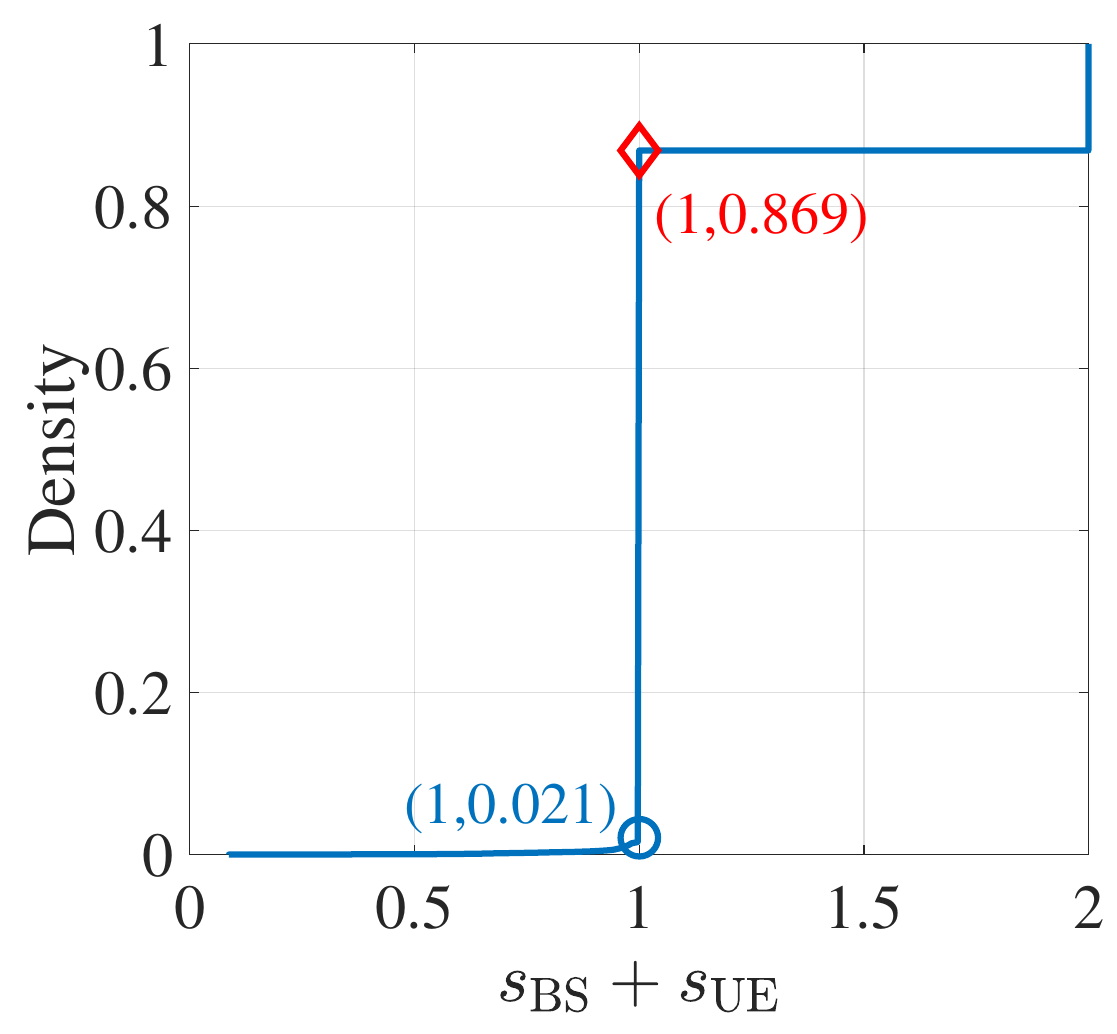}}}
\subfloat[]{
{\includegraphics[width=4.3cm]{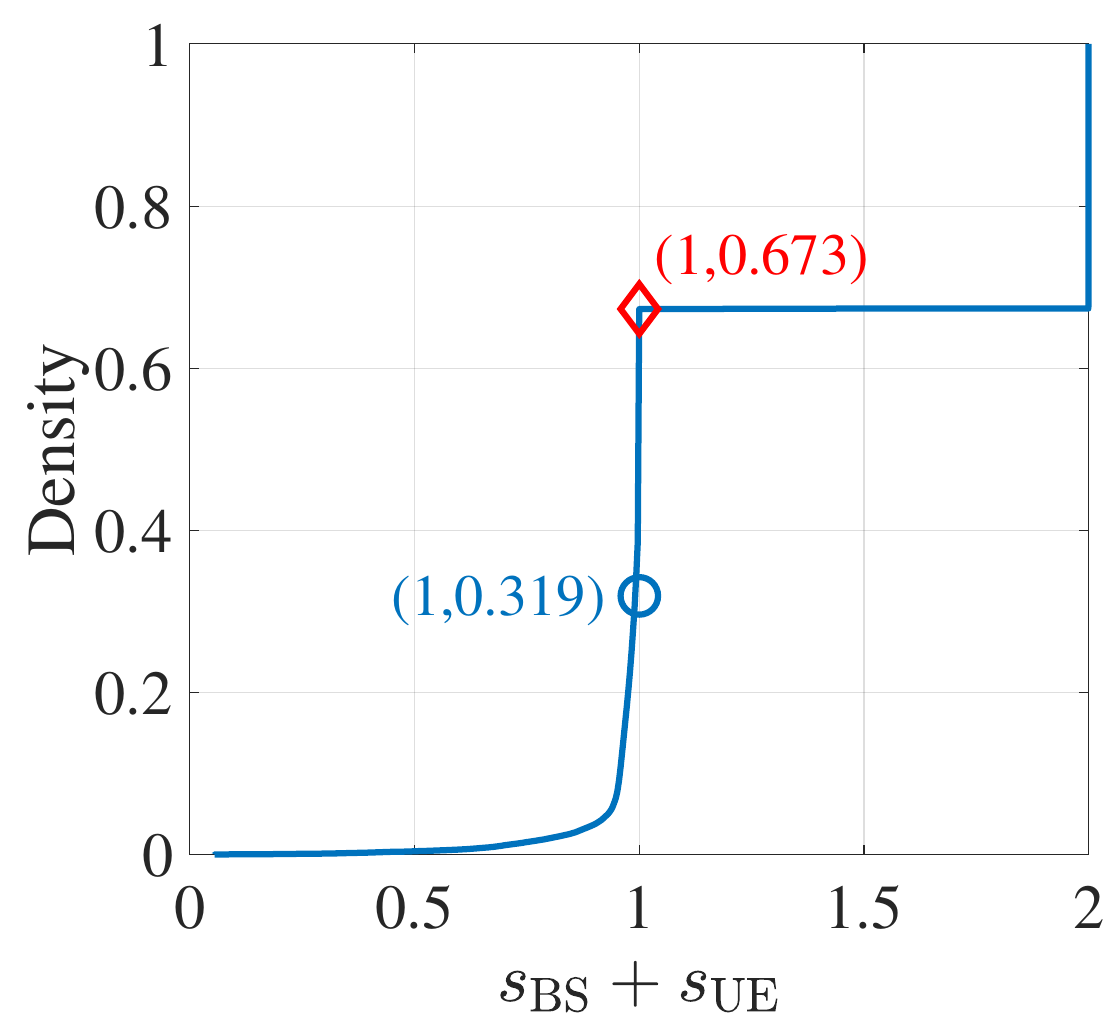}}
}
\caption{CDFs of the sum of scaling factors at the BS and UE sides in (a) UMi and (b) InH scenarios.}
\label{fig_NF_s_TRP_s_UE}
\end{figure}

To determine $s_{\mathrm{UE}}$, we analyze the statistical distribution of the sum of the scaling factors at the BS and UE, defined as: $s_{\mathrm{sum}} = s_{\mathrm{BS}} + s_{\mathrm{UE}}$. As shown in Fig.~\ref{fig_NF_s_TRP_s_UE}, $s_{\rm{sum}}$ predominantly takes values of 1 or 2. A value of 2 corresponds to purely specular reflection paths, where $s_{\mathrm{BS}} = s_{\mathrm{UE}} = 1$. A value of 1 suggests only one non-specular interaction, in which case $s_{\mathrm{UE}} = 1 - s_{\mathrm{BS}}$. Rare cases where $s_{\mathrm{sum}} < 1$ imply multiple non-specular interactions, which contribute little power due to high path loss and can be safely ignored. This structure simplifies the UE-side spherical-wave source distance calculation: once $s_{\mathrm{BS},n}$ is known, $s_{\mathrm{UE},n}$ can be derived as:
\begin{equation}
    {s_{{\rm{UE}},n}} = \left\{ {\begin{array}{*{20}{c}}
1&{n \le N_{\rm{spec}}},\\
{1 - {s_{{\rm{BS}},n}}}&{n > N_{\rm{spec}}}.
\end{array}} \right.
\end{equation}

Finally, using the generated distances $d_{1,n,m}$ and $d_{2,n,m}$, the direction vectors from each TX and RX element to the spherical-wave source can be computed as $\vec{r}_{{tx},n,m,s} = d_{1,n,m} \cdot \hat{r}_{{tx},n,m} - \bar{d}_{{tx},s}$ and $\vec{r}_{{rx},n,m,u} = d_{2,n,m} \cdot \hat{r}_{{rx},n,m} - \bar{d}_{{rx},u}$, which serve as the basis for determining the near-field array phase and angular-domain parameters.

\section{Spatial Non-Stationarity Feature}

The SNS feature characterizes element-wise power variations independently at the BS and UE sides.

\subsection{Spatial Non-Stationarity Feature at the BS Side}

To characterize the SNS feature at the BS side, two modeling approaches are supported: a stochastic-based method and a physical blocker-based method. Among them, the stochastic-based model is recommended as the primary option for simulation. The following guidelines help determine the appropriate model:
\begin{itemize}
\item Stochastic-based model: Suitable for simulations that aim to capture the effects of incomplete scattering, or in scenarios where computationally efficient SNS modeling is preferred.
\item Physical blocker-based model: Recommended for scenarios that involve partial blockage, or in scenarios where higher physical realism and consistency are required.
\end{itemize}

\subsubsection{Stochastic-Based Model}
In the stochastic-based model, the concept of VRs is adopted, wherein each cluster is visible only to a subset of the antenna array. Outside this region, the cluster’s power is assumed to attenuate gradually. It is worth noting that this approach is modeled at the cluster level, i.e., the power attenuation factor is uniform across all rays within the same cluster, such that $\alpha_{s,n,m} = \alpha_{s,n}$ for all $m$. The CIR for the NLOS component is expressed as:
\begin{equation}
\begin{aligned}
        H_{u,s}^{\rm{NLOS}}\left( {\tau ,t} \right) = \underbrace {\sum\limits_{n \notin {{\cal N}_{{\rm{SNS}}}}} {\sum\limits_{m = 1}^M {H_{u,s,n,m}^{\rm{NLOS}}\delta \left( {\tau  - {\tau _{n,m}}} \right)} } }_{{\rm{SS \ clusters}}} \\+ \underbrace {\sum\limits_{n \in {{\cal N}_{{\rm{SNS}}}}} {\sqrt {{\alpha _{s,n}}} \sum\limits_{m = 1}^M {H_{u,s,n,m}^{\rm{NLOS}}\delta \left( {\tau  - {\tau _{n,m}}} \right)} } }_{{\rm{SNS \ clusters}}},
\end{aligned}
\end{equation}
where $\mathcal{N}_{\mathrm{SNS}}$ denotes the set of SNS clusters, for which the synthesized CIR includes an attenuation factor $\alpha_{s,n}$. In contrast, the SS clusters remain unaffected, i.e., $\alpha_{s,n} = 1$. For simplicity, the delay parameter is denoted as $\tau_{n,m}$, representing the propagation delay associated with each ray. In the case of the LOS condition, the LOS path is considered as an additional cluster. Therefore, the stochastic-based model requires the identification of SNS clusters, their corresponding VRs, and the associated power attenuation factors.

\paragraph{Identification of SNS Clusters}

\begin{figure}[htbp]
\centering
\subfloat[]
{\includegraphics[width=4.3cm]{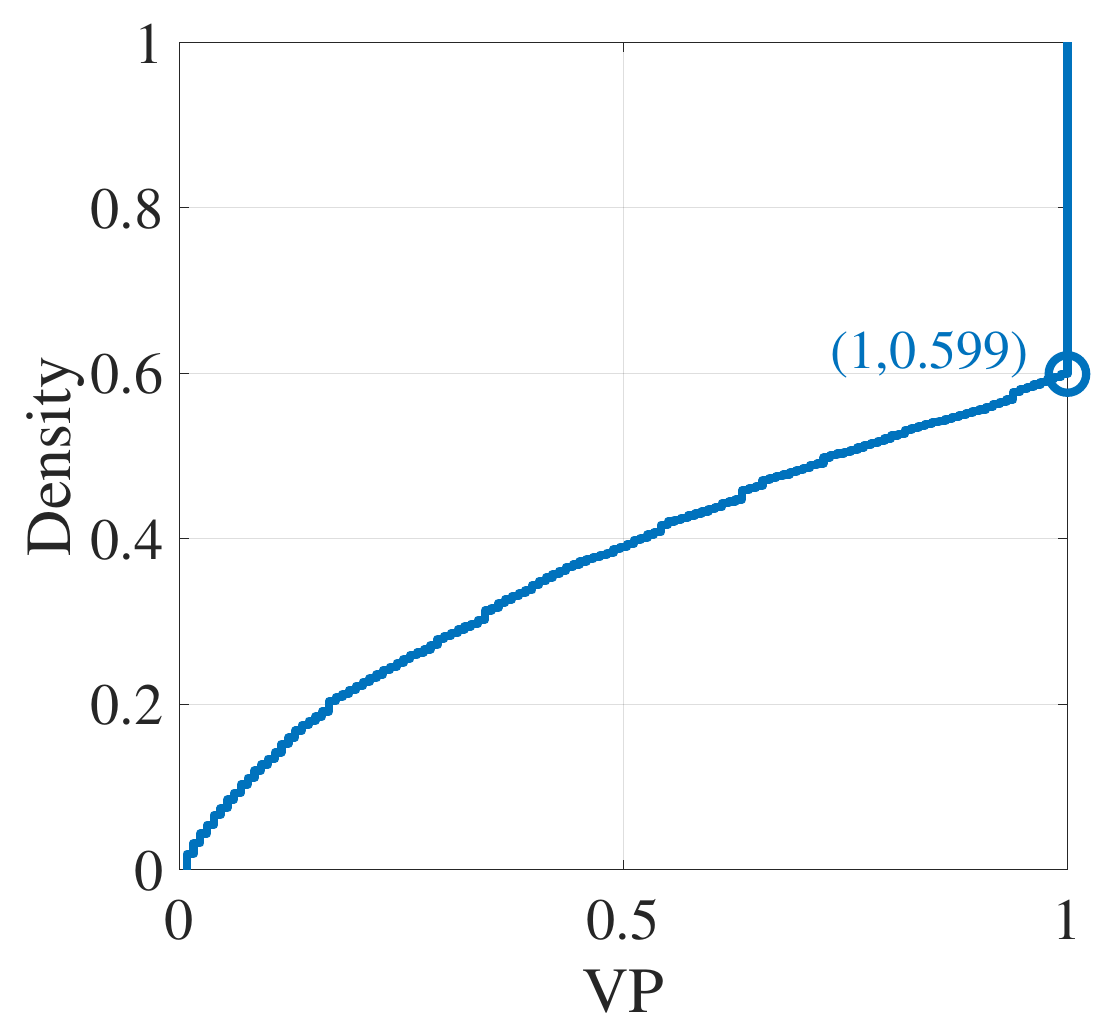}
\label{fig_SNS_Pr_a}}
\hfill
\subfloat[]
{\includegraphics[width=4.3cm]{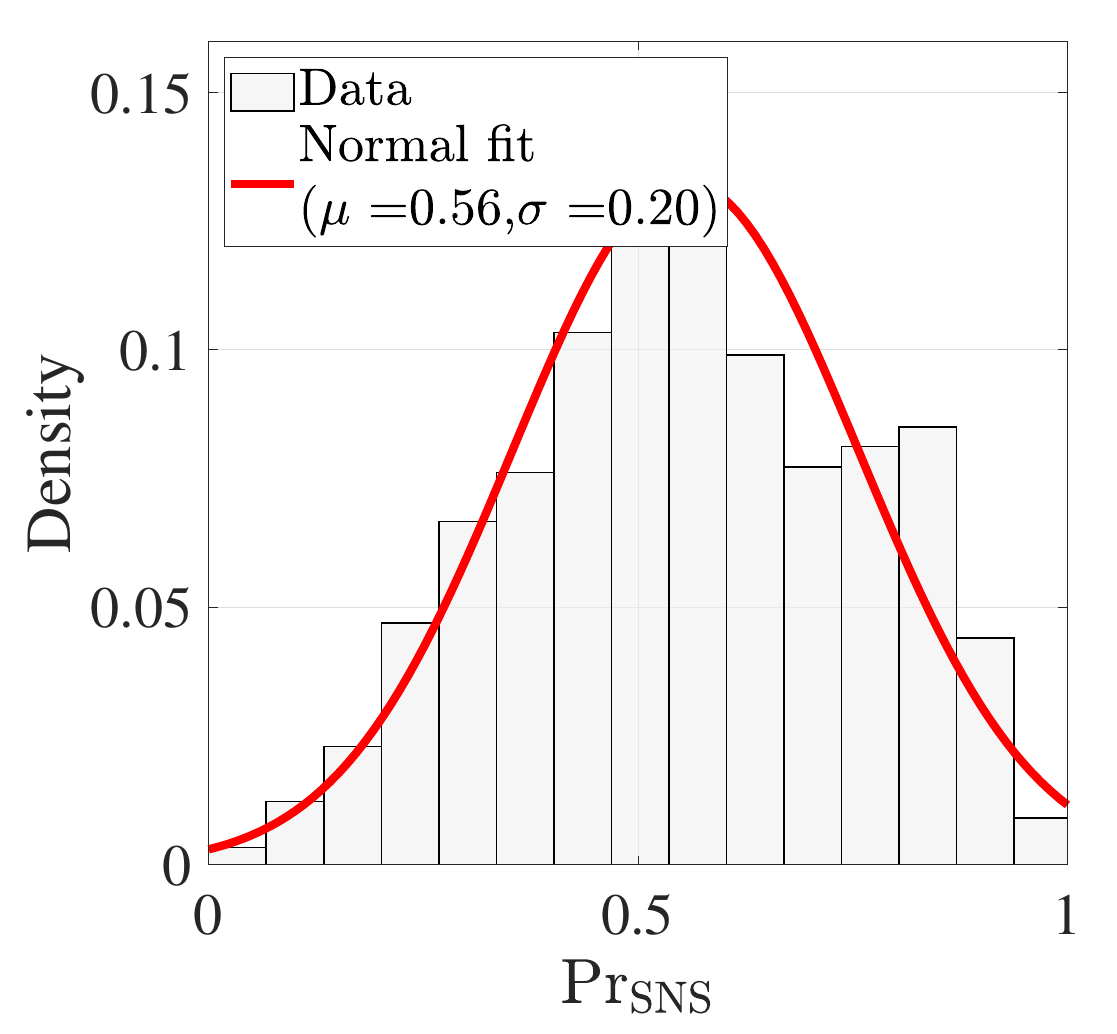}
\label{fig_SNS_Pr_b}}
\caption{(a) Cluster VP; (b) SNS probability of UEs.}
\vspace{-0.1cm}
\label{fig_SNS_Pr}
\end{figure}

Although the aperture of XL-MIMO arrays increases significantly, not all clusters exhibit SNS—some SS clusters may still exist. The VP is defined as the ratio of the number of antenna elements where the cluster is visible to the total number of antenna elements \cite{3GPP_huawei_fugang}, i.e.,
\begin{equation}
    V_n = \dfrac{S_{n}^{\rm{vis}}}{S_{\rm{all}}},
    \label{equ_VP}
\end{equation}
where $S_{n}^{\rm{vis}}$ is the number of visible antenna elements for the $n$-th cluster, and $S_{\rm{all}}$ is the total number of elements in the XL-MIMO array. Ray-tracing simulation results in the UMa scenario are presented in Fig.~\ref{fig_SNS_Pr_a}, which show that approximately 40.1\% of clusters have a VP of 1, indicating full visibility across the array and thus classification as SS clusters.

Accordingly, for each UE, the SNS probability ${\rm{Pr}}_{\mathrm{SNS}}$ is defined as the proportion of SNS clusters relative to the total number of clusters. By statistically analyzing the SNS probabilities of all ray-tracing simulated UEs, it is observed that the distribution of ${\rm{Pr}}_{\mathrm{SNS}}$ follows a normal distribution, as illustrated in Fig.~\ref{fig_SNS_Pr_b}. Additional scenario results such as UMi, InH, InF, rural macro (RMa), and suburban macro (SMa) are summarized in Table~\ref{Tab_SNS_VPVR}.

During model implementation, a random value drawn from the fitted normal distribution (bounded within $[0, 1]$) can be used to represent the ${\rm{Pr}}_{\mathrm{SNS}}$ of each UE. Then, for each cluster associated with the UE, a uniform random variable $x \sim {\cal{U}}(0,1)$ is generated. If $x < {\rm{Pr}}_{\mathrm{SNS}}$, the cluster is classified as SNS; otherwise, it is considered SS.

\paragraph{Visibility Region}

\begin{figure}[htbp]
\centering
{\includegraphics[width=8cm]{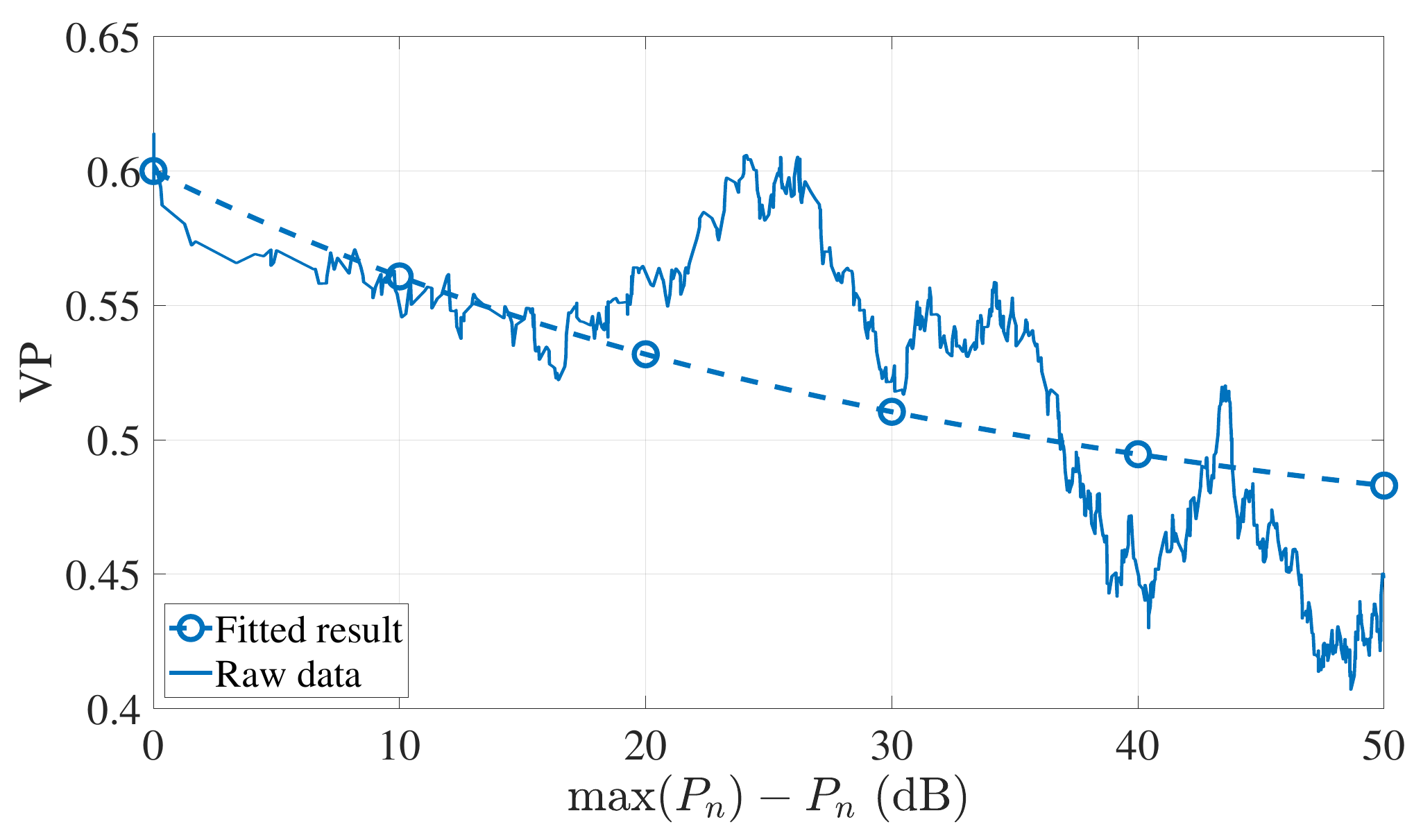}}
\caption{Relationship between cluster VP and power in UMa scenario.}
\label{fig_SNS_VP}
\end{figure}

For SNS clusters, the VR is modeled as a rectangular area with dimensions $a \cdot b$, where $a$ and $b$ denote the width and height of the VR, respectively. The width $a$ is randomly generated from a uniform distribution as $a \sim {\cal{U}}(V_n \cdot W,\, W)$. The height $b$ is then computed as $b = V_n \cdot H \cdot W / a$, where $H$ and $W$ denote the vertical and horizontal dimensions of the antenna array. This construction guarantees that the total area of the VR is consistent with the definition of VP in Eq.~(\ref{equ_VP}). In addition, one of the four corners of the antenna array is randomly selected as the reference corner from which the rectangular VR originates. Ray-tracing simulation results show a correlation between the VP of a cluster and its power \cite{3GPP_huawei_wuhan}, as illustrated in Fig.~\ref{fig_SNS_VP}. This relationship is modeled as:
\begin{equation}
    {{V}}_n = A \cdot \exp \left(- \dfrac{ \max(P_n)-P_n}{R}\right)+B+\xi,
\end{equation}
where $A$ and $R$ are parameters of the exponential decay model, $B$ is the lower bound of $V_n$, and $\xi \sim \mathcal{N}(0, \sigma^2)$. $P_n$ represents the power of the $n$-th cluster in dB. The fitting parameters for the remaining scenarios are summarized in Table~\ref{Tab_SNS_VPVR}.

\paragraph{Power Attenuation Factor}

\begin{figure}[htbp]
\centering
{\includegraphics[width=7.6cm]{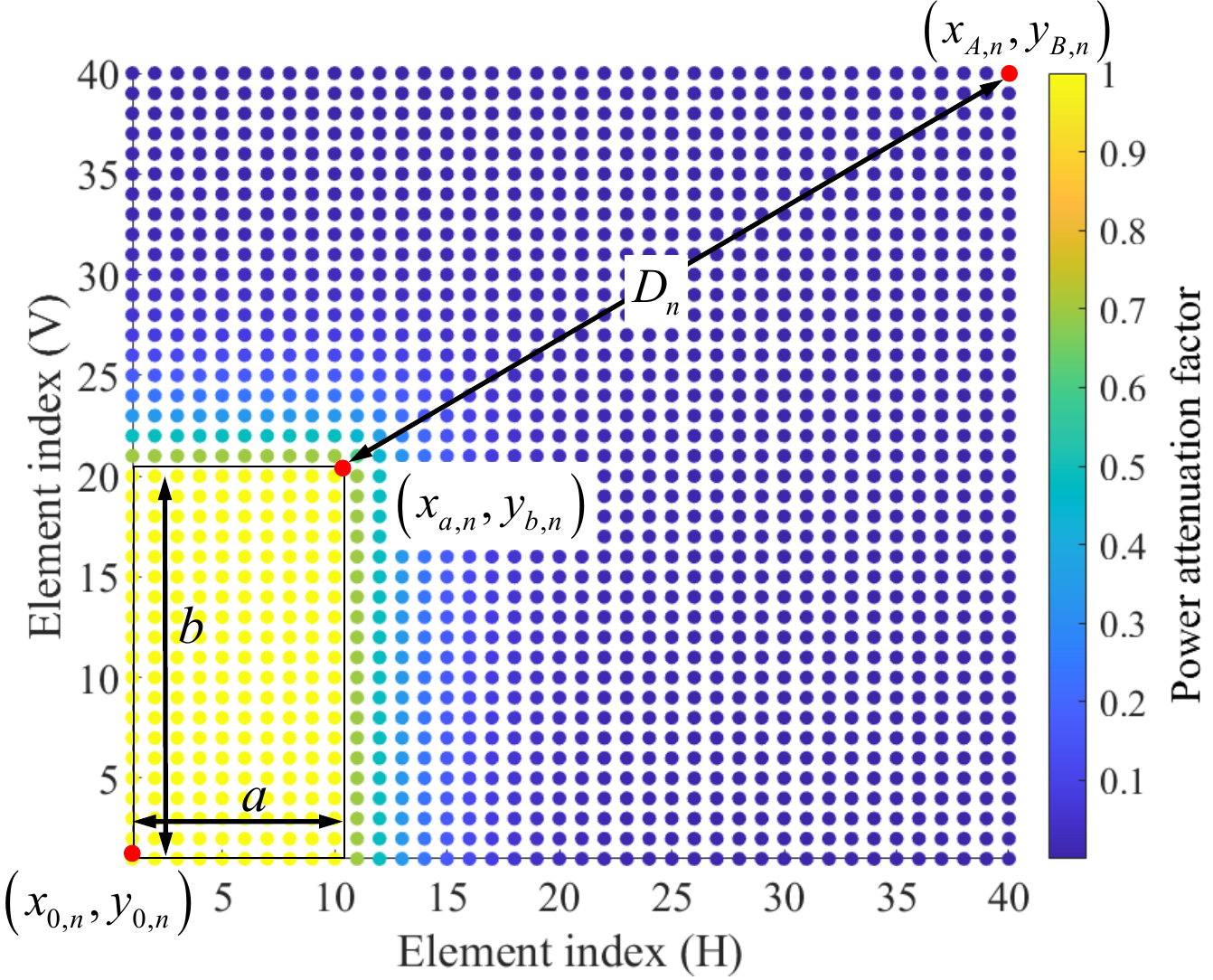}}
\caption{Illustration of the power attenuation factor across antenna elements.}
\vspace{-0.2cm}
\label{fig_SNS_PAF}
\end{figure}

The power attenuation factor $\alpha_{s,n}$ applied to each antenna element $s$ can be categorized into two cases:
\begin{itemize}
    \item For SS clusters, each antenna element experiences equal power without attenuation, i.e., $\alpha_{s,n} = 1$.
    \item For SNS clusters, the power attenuation factor \(\alpha_{s,n}\) at the $s$-th antenna element is computed as \cite{3GPP_huawei_wuhan}:
        \begin{equation}
        {\alpha _{s,n}} = \left\{ {\begin{array}{*{20}{c}}
{1,}&(x_s, y_s) \in \mathcal{R} ,\\
{{\exp{\left( { - C \cdot \dfrac{{ {\tilde d_{s,n}}}}{{{D_n}}}} \right)}},}&(x_s, y_s) \notin \mathcal{R},\\
\end{array}} \right.
\label{equ_SNS_PAF}
    \end{equation}
    where $\mathcal{R} = \{ |x_s - x_{0,n}| \le a,\, |y_s - y_{0,n}| \le b \}$ denote the VR of the \(n\)-th cluster, with $(x_s, y_s)$ representing the coordinates of antenna element $s$, and $(x_{0,n}, y_{0,n})$ being the reference corner of the VR. The diagonally opposite corner of the VR is denoted as $(x_{a,n}, y_{b,n})$, while $(x_{A,n}, y_{B,n})$ corresponds to coordinate of the other corner of the antenna array on the diagonal with reference corner $(x_{0,n}, y_{0,n})$. The distance $\tilde d_{s,n}$ from element $s$ to the nearest point on the VR boundary is given by $x_s - x_{a,n}$ when $|x_s - x_{0,n}| > a$ and $|y_s - y_{0,n}| \le b$, by $y_s - y_{b,n}$ when $|x_s - x_{0,n}| \le a$ and $|y_s - y_{0,n}| > b$, and by $\sqrt{(x_s - x_{a,n})^2 + (y_s - y_{b,n})^2}$ otherwise. ${D_n} = \sqrt {{{\left( {{x_{A,n}} - {x_{a,n}}} \right)}^2} + {{\left( {{y_{B,n}} - {y_{b,n}}} \right)}^2}}$. The roll-off factor $C$ controls the transition sharpness between the visible and invisible regions and is typically set to 13 for all scenarios defined by 3GPP. As shown in Fig.~\ref{fig_SNS_PAF}, elements within the VR experience no power attenuation. Outside the VR, the attenuation factor gradually decreases with distance from the VR, reflecting a smooth transition in cluster power.
\end{itemize}

\begin{table}
\renewcommand\arraystretch{1.2}
    \centering
    \caption{Parameters of the stochastic-based SNS model}
    \begin{tabular}{m{1cm}<{\centering}|m{0.3cm}<{\centering}|m{0.6cm}<{\centering}|m{0.6cm}<{\centering}|m{0.6cm}<{\centering}|m{0.6cm}<{\centering}|m{0.6cm}<{\centering}|m{0.6cm}<{\centering}}
    \hline
       \multicolumn{2}{c|}{Scenario}  & UMa & UMi & InH & InF & RMa & SMa\\
         \hline
        \multirow{2}{*}{${\rm{Pr}}_{\mathrm{SNS}}$} & \(\mu\) & 0.56 &0.49 & 0.31  & 0.32 & 0.14 & 0.24 \\
        \cline{2-8}
       & \(\sigma\) & 0.20 &0.18 & 0.08 & 0.06 & 0.08 & 0.07  \\
       \hline
        \multirow{4}{*}{\(V_n\)} & \(A\) & 0.15 &0.12 & 0   & 0 & 0.16 & 0.06\\
        \cline{2-8}
       & \(B\) & 0.45 &0.48 & 0.60 &0.57 & 0.74 & 0.56   \\ \cline{2-8}
        & \(R\) & 33 & 50 & - & - & 60 & 23  \\ \cline{2-8}
         & \(\sigma^2\) &0.0015 &0.001 & 0.0011 & 0.002 & 0.0016 & 0.0013  \\ 
         \hline
    \end{tabular}
    \label{Tab_SNS_VPVR}
\end{table}

\subsubsection{Physical Blocker-Based Model}

Since partial blockage is a key physical cause of SNS, a natural alternative approach is to adopt the blockage model specified in Section 7.6.4.2 of 3GPP TR 38.901 \cite{C3_3GPP}. The implementation primarily follows Blockage Model B, where multiple blockers are deployed. Their size, position, density, and movement patterns are determined by simulation assumptions, allowing flexible configuration of various blockage scenarios depending on the study’s objectives. To specifically reflect SNS effects, four new blocker types are introduced—billboards, street lamps, building edges, and pillars.

The antenna element-wise power variation caused by each blocker to each ray for each element is modeled using a simple knife-edge diffraction model and is given by:
\begin{equation}
\begin{aligned}
    L_{{\rm{dB}},u,s,1}/&L_{{\rm{dB}},s,n,m} = \\
    &- 20{\log _{10}}\left( {1 - \left( {{F_{h1}} + {F_{h2}}} \right)\left( {{F_{w1}} + {F_{w2}}} \right)} \right),
\end{aligned}
\label{equ_SNS_Blockage}
\end{equation}
where $F_{h1}$, $F_{h2}$, $F_{w1}$, and $F_{w2}$ represent the diffraction loss contributions from the four edges, as defined in Equation (7.6-30) of TR 38.901. By converting $L_{{\rm{dB}},u,s,1}$ or $L_{{\rm{dB}},s,n,m}$ to a linear scale, the corresponding element-wise power attenuation factor can be obtained. Compared to the original Blockage Model B, two key modifications are introduced:
\begin{itemize}
    \item Element-wise, ray-specific orientation: For each ray and each BS-side antenna element, the blocker screen is rotated around its center such that the departure direction of the corresponding path is always perpendicular to the screen. This rotation must be computed individually for each ray and antenna element. Moreover, the projected distances to the four blocker edges must be evaluated per element, rather than relative to the overall TX/RX location.
    \item Simplified modeling for large blockers (e.g., building edges): Buildings are usually too large for all four edges to contribute to diffraction. Therefore, only a single edge is typically considered in the knife-edge attenuation model. For the remaining edges, projected distances are treated as infinite, with their diffraction factors equal to 0.5. Under this assumption, the Eq.~(\ref{equ_SNS_Blockage}) simplifies as \(- 20{\log _{10}}\left( {0.5 - {F}{_{{h_1}|{h_2}|{w_1}|{w_2}}}} \right)\).
\end{itemize}

\subsection{Spatial Non-Stationarity Feature at the UE Side}

Taking into account practical device usage, element-wise power variation at the UE side can arise from partial obstruction of the antenna array by the user’s head or hands. According to smartphone usage statistics, approximately 90\% of UEs experience such obstructions. These scenarios correspond to common usage scenarios: one-hand grip during social media, productivity, shopping, and browsing (58\%), two-hand grip for video streaming and gaming (29\%), and one-hand with head obstruction during voice calls (13\%).

A straightforward approach to model this effect is to assign fixed attenuation values to each antenna element, reflecting the element-wise power loss due to obstruction. However, these attenuation levels vary with frequency bands, due to differences in penetration loss and antenna design. For example, antennas below 1 GHz often use the entire device chassis as a radiator, since the chassis length is roughly a quarter-wavelength. In contrast, higher-frequency antennas ($\ge$ 1 GHz) typically excite only part of the chassis \cite{3GPP_Nokia_wuhan}. Furthermore, attenuation caused by hand and head blockage is frequency-dependent as shown in measurement and simulation studies \cite{SNS_UE}. To capture this frequency dependency, power attenuation values are defined separately for different frequency ranges (i.e., below 1 GHz, 1–8.4 GHz, and 14.5–15.5 GHz). These values are derived from a combination of measurements, full-wave electromagnetic simulations, and ray-tracing analyses conducted under various user-related conditions, as summarized in Table~7.6.14.2-2 of \cite{3GPP_CR}. By converting these fixed attenuation values to a linear scale, the element-wise power attenuation factors \(\beta_{u}\) at the UE side are obtained.

\section{XL-MIMO Channel Model Implementation and Performance Evaluation}

\subsection{Channel Model Implementation}

\begin{figure*}[htbp]
\centering
\includegraphics[width=16.5cm]{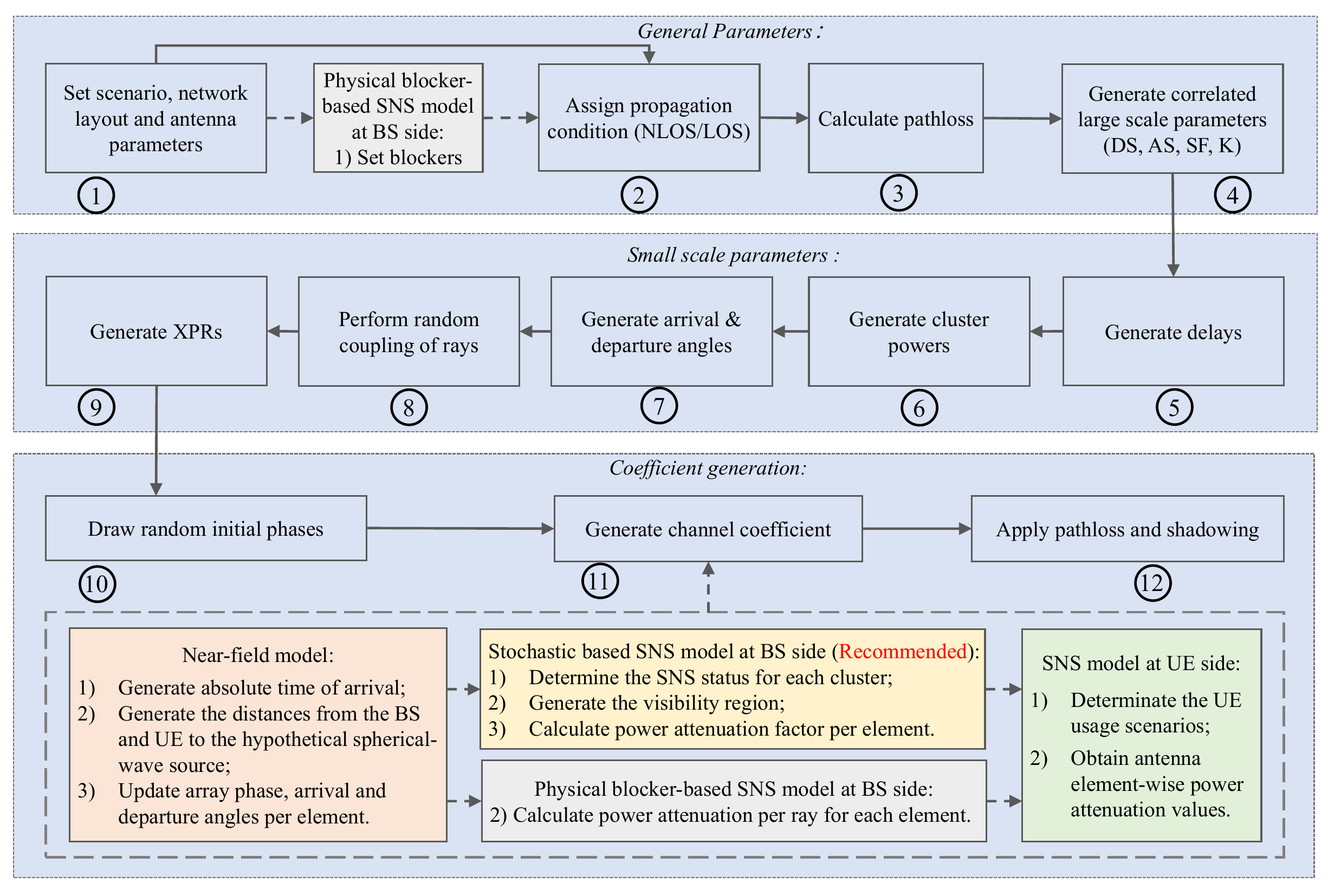}
\caption{Channel coefficient generation procedure: evolution from MIMO to XL-MIMO.}
\vspace{-0.5cm}
\label{Fig_flowchart}
\end{figure*}

The 3GPP TR 38.901 defines a widely adopted channel model for 5G wireless systems \cite{C3_3GPP}. The overall channel coefficient generation procedure is illustrated by the solid lines in Fig.~\ref{Fig_flowchart}, which is organized into three main modules comprising twelve steps. In the general parameters module, large-scale parameters such as shadow fading, delay spread, and angular spread are generated based on statistical distributions. In the small-scale parameters module, delays, powers, angles of arrival and departure, and XPRs are generated following specified procedures. In the coefficient generation module, CIRs are synthesized using Eqs.~(\ref{equ_CIR_LOS})-(\ref{equ_H_NLOS_5G}), followed by the application of path loss and shadowing.

To incorporate near-field propagation, three extra steps are added in Step 11. Specifically,
\begin{itemize}
\item Generate absolute time of arrival: Combine the relative delays from Step 5, the 3D LOS propagation delay, and the excess NLOS delay.
\item Generate the distances from the BS and UE to the spherical-wave source associated with clusters: For both specular and non-specular reflection clusters, generate the corresponding scaling factors, and then multiply them by the absolute delay and the speed of light to obtain the BS-to-source and UE-to-source distances.
\item Update array phase, arrival and departure angles per element: Based on the computed distances, determine the direction vector of each ray for each TX/RX antenna element, and calculate the element-wise array phase and angular parameters to update the channel coefficients.
\end{itemize}

For SNS model implementation, the process is divided into two parts: one at the BS side and the other at the UE side. At the BS side, two alternative approaches are supported. The recommended stochastic-based SNS model is integrated into Step 11, with the following procedure:
\begin{itemize}
\item Determine the SNS status for each cluster: Based on the SNS probability of each UE, determine whether each cluster is SNS.
\item Generate the VR: For each SNS cluster, compute its VP based on its power, and randomly define a rectangular VR that satisfies the VP requirement.
\item Calculate power attenuation factor per element: For SS clusters, the attenuation factor is set to 1 across the array. For SNS clusters, the factor is 1 inside the VR and gradually decreases outside. The resulting values are applied in coefficient generation.
\end{itemize}

For the physical blocker-based SNS model, the steps are as follows:
\begin{itemize}
\item Set blockers: Deploy various types of blockers after Step 1, according to the simulation setup.
\item Calculate power attenuation per ray for each element: Use the knife-edge diffraction model to compute the element-wise attenuation for each ray and apply it in coefficient generation.
\end{itemize}

At the UE side, SNS modeling is addressed in Step 11, with the following procedure:
\begin{itemize}
\item Determine the UE usage scenario: Randomly select one of the four scenarios—one-hand grip, dual-hand grip, head and one-hand grip, or free space—based on their occurrence probabilities.
\item Obtain antenna element-wise power attenuation values: According to the usage scenario and operating frequency, retrieve per-element attenuation values and apply them during coefficient generation.
\end{itemize}

\vspace{-0.5cm}
\subsection{Performance Evaluation}

To validate the feasibility of the XL-MIMO near-field propagation and SNS channel modeling framework, we simulate two representative scenarios: UMi and InH. The simulation setup is summarized in Table~\ref{Tab_Performance}, with 1,000 UEs randomly deployed per simulation run.

\begin{table}
\renewcommand\arraystretch{1.2}
    \centering
    \caption{Simulation assumptions for performance evaluation}
        \begin{tabular}{m{2.5cm}<{\centering}|m{2.2cm}<{\centering}|m{2cm}<{\centering}}
    \hline
      Parameters & \multicolumn{2}{c}{Values}\\
          \hline
        Scenarios & UMi & InH  \\
        \hline
        Carrier frequency & \multicolumn{2}{c}{7 GHz }    \\
        \hline
       BS antenna height  & 10 m &3 m  \\
       \hline
       Signal-to-noise ratio (SNR) & \multicolumn{2}{c}{10 dB}  \\
       \hline
       BS antenna configuration & \multicolumn{2}{c}{\makecell*[c]{64×16 dual-polarized UPA, \\half-wavelength spacing, directional pattern}}   \\
       \hline
       UE antenna configuration &\multicolumn{2}{c}{\makecell*[c]{8 antennas distributed on a handheld device,\\ dual-polarization \cite{3GPP_CR}}}   \\
       \hline
       UE antenna pattern & \multicolumn{2}{c}{Isotropic}   \\
       \hline
       UE location &LOS scenario with outdoor UE at 1.5~m height &LOS scenario with indoor UE at 1 m height  \\
       \hline
       \multicolumn{3}{c}{Radius for uniform UE deployment centered at the BS} \\
       \hline
       Near-field simulation &20, 50, and 100 m & 2, 5, and 10 m \\
       \hline
       SNS simulation & 100 m & 10 m \\
        \hline
    \end{tabular}
    \vspace{-0.5cm}
    \label{Tab_Performance}
\end{table}

\begin{figure}[htbp]
\centering
\subfloat[]
{\includegraphics[width=8.3cm]{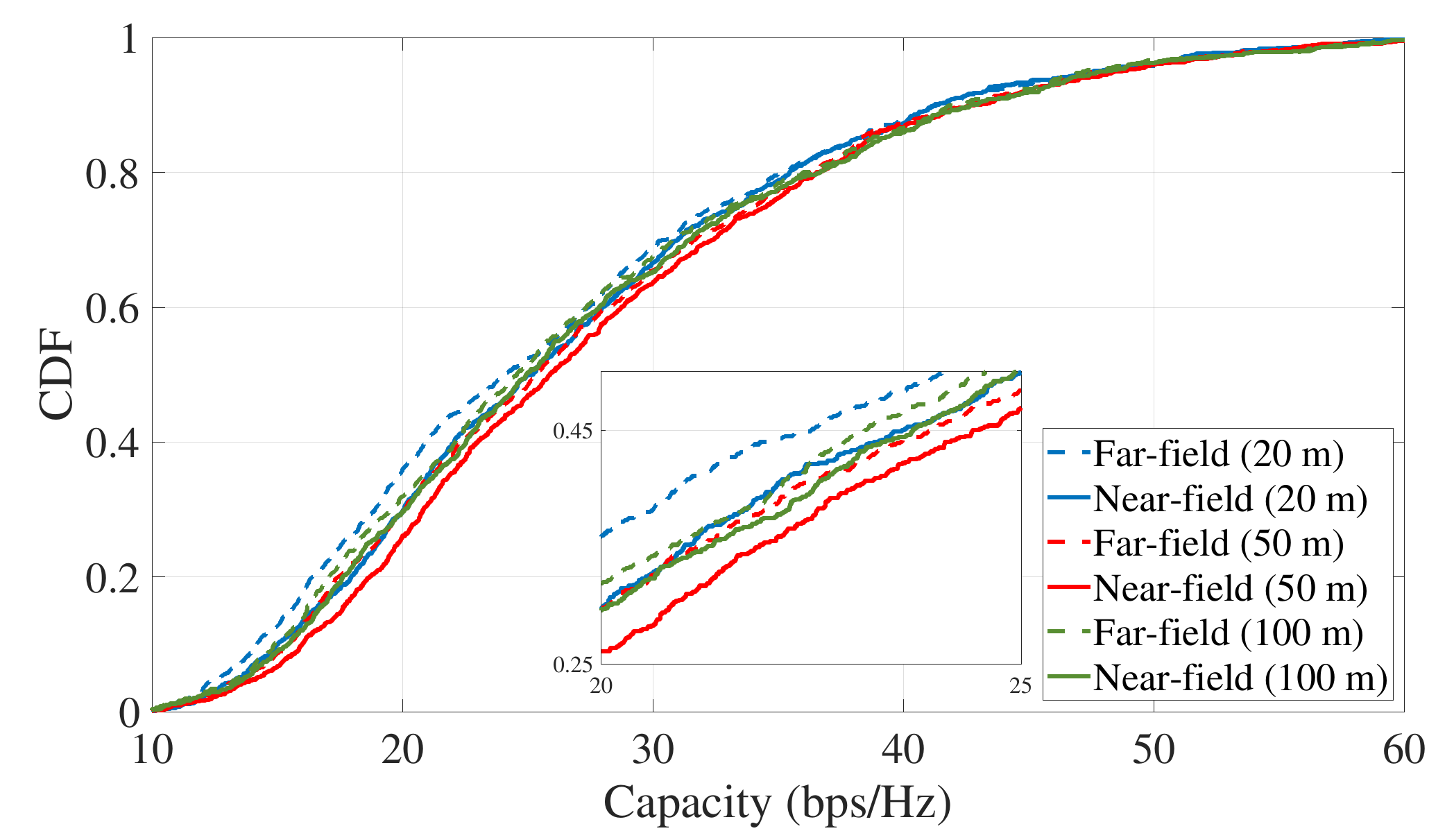}
\label{fig_NF_Capacity_a}}
\hfill
\subfloat[]
{\includegraphics[width=8.3cm]{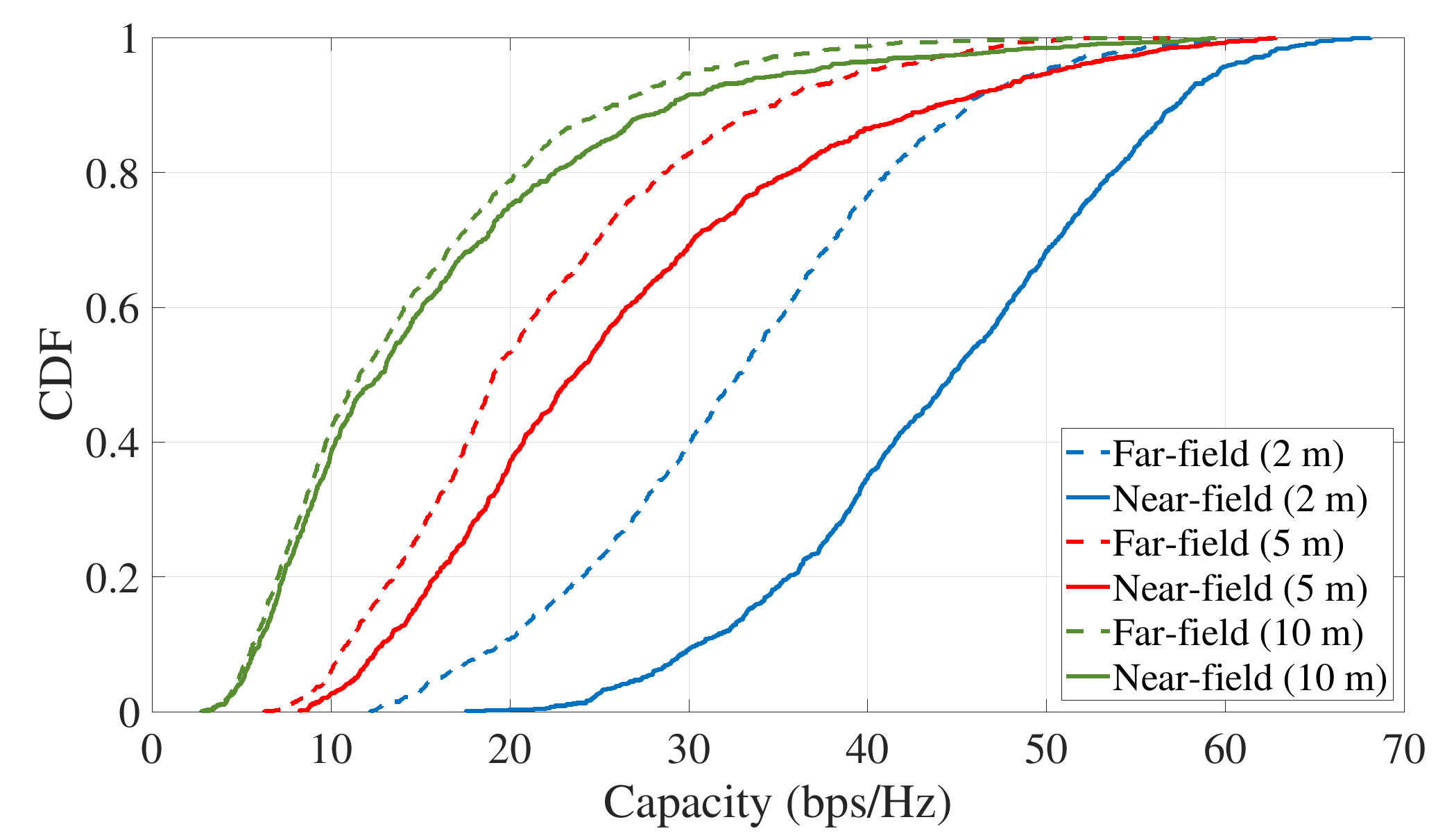}
\label{fig_NF_Capacity_b}}
\caption{Channel capacity results for (a) UMi and (b) InH scenarios.}
\label{fig_NF_Capacity}
\end{figure}

Fig.~\ref{fig_NF_Capacity} compares the channel capacity under the near-field and far-field models. As expected, the near-field model provides consistent capacity improvement. In the UMi scenario, the gain is relatively modest, with average increases of  0.70, 0.59, and 0.44 bps/Hz for UE deployment radii of 20, 50, and 100 m, respectively. This is primarily due to the larger 3D distances between BS and UE and the horizontal distance constraint of at least 10 m, which limits near-field advantages. In contrast, the shorter BS–UE distances in the InH scenario lead to more substantial capacity gains: 11.60, 4.75, and 1.46 bps/Hz for radii of 2, 5, and 10 m, respectively. These results also indicate that capacity gains become more pronounced as the UE is placed closer to the BS.

\begin{figure}[htbp]
\centering
{\includegraphics[width=8.3cm]{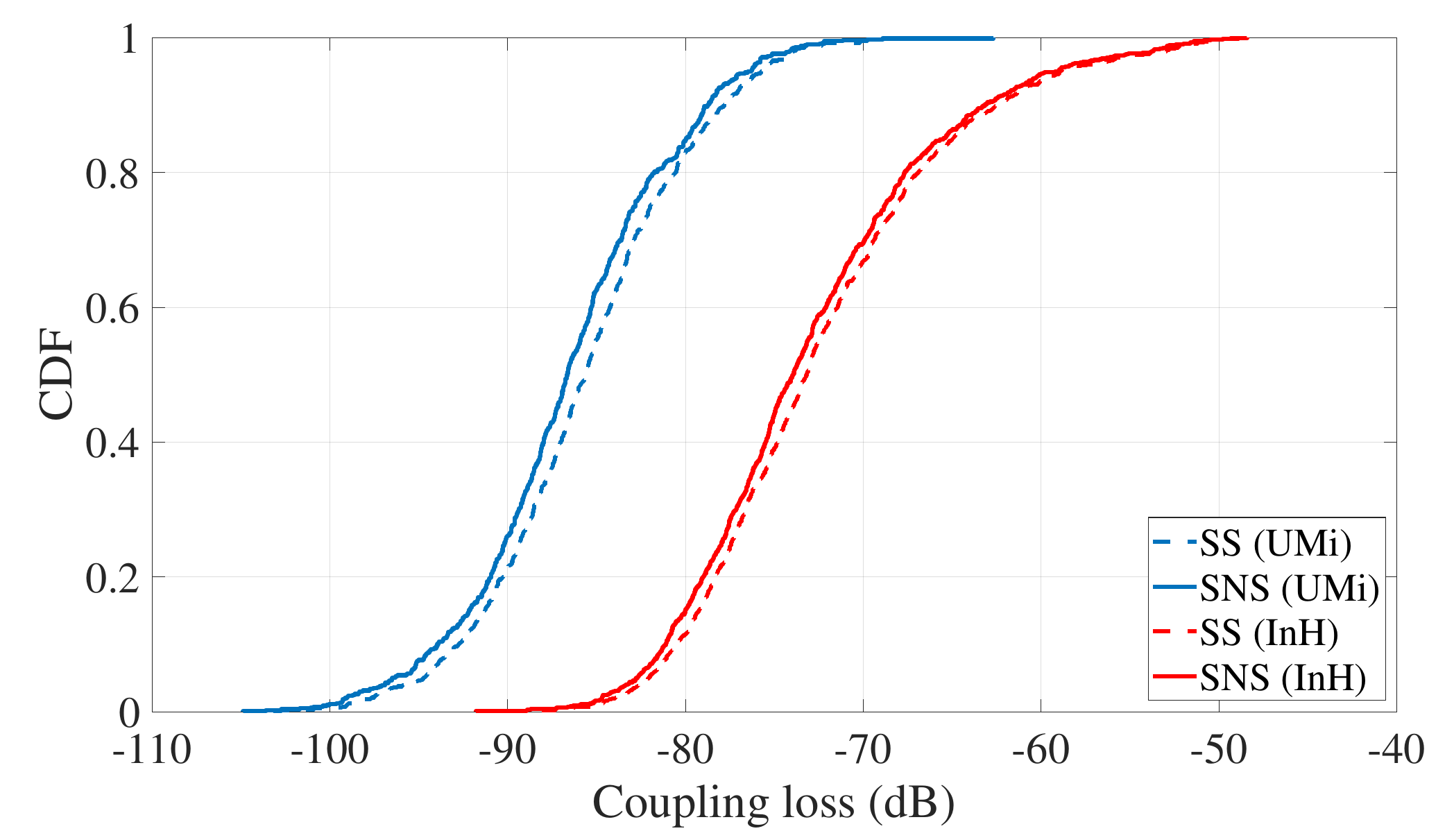}}
\caption{Coupling loss results for UMi and InH scenarios.}
\label{fig_SNS_ACG}
\end{figure}

To assess the impact of SNS at the BS side, we evaluate the coupling loss, which is defined as the channel gain between the UE and its serving BS. This channel gain is calculated by averaging the power across all sublinks in the MIMO channel matrix and incorporates both path loss and shadowing effects \cite{Quadriga}.  The recommended stochastic-based SNS model is applied, and the simulation results for the UMi and InH are shown in Fig.~\ref{fig_SNS_ACG}. As observed, enabling the SNS feature reduces the coupling loss, implying stronger spatial fading, with reductions of 0.91 dB in the UMi scenario and 0.67 dB in the InH scenario. The InH scenario exhibits a smaller reduction due to its lower SNS probability and higher VP compared to UMi.

\vspace{-0.1cm}
\section{Conclusion}

This paper presents the channel modeling framework adopted by 3GPP for capturing near-field propagation and SNS for XL-MIMO systems. The framework introduces key parameter extensions, including distance parameters from the BS and UE to the spherical-wave source associated with clusters for characterizing near-field effects, and element-wise power attenuation factors to capture SNS at both BS and UE sides. For near-field modeling, the distances associated with specular reflection clusters are set equal to the actual propagation path length, while those for non-specular clusters are computed by multiplying the propagation path length by a scaling factor sampled from a Beta distribution. For SNS modeling at the BS side, two approaches are supported. The stochastic-based approach introduces the concept of VP to define a cluster-specific VR, where antenna elements within the VR experience no attenuation, and those outside experience a smooth attenuation. Alternatively, the physical blocker-based approach builds on the already existing 3GPP Blockage Model B to compute per-element loss. At the UE side, SNS is modeled through fixed per-element attenuation values. Finally, a simulation framework incorporating these features has been developed. Performance evaluations show that the near-field model reveals higher channel capacity compared to the far-field model, while incorporating SNS leads to more pronounced signal fading.






\bibliographystyle{IEEEtran}
\bibliography{paper}
 

 


\vfill

\end{document}